\documentclass[12pt,a4paper]{article}

\usepackage{changepage}
\usepackage[utf8]{inputenc}
\usepackage[T1]{fontenc}

\usepackage[a4paper,top=2cm,bottom=2cm,left=3cm,right=3cm,marginparwidth=1.75cm]{geometry}

\usepackage[affil-it]{authblk} 

\usepackage[section]{placeins}

\usepackage{amsmath}
\usepackage{graphicx}
\usepackage{tikz}

\usepackage[colorinlistoftodos]{todonotes}
\usepackage[colorlinks=true, allcolors=blue]{hyperref}
\usepackage{mathptmx}
\usepackage{soul,color}
\usepackage{amsmath}
\usepackage{mathrsfs}
\usepackage{amsmath,empheq}
\usepackage{amsfonts}
\usepackage[font=bf]{caption}
\usepackage{comment}
\usepackage{titlesec}

\titleformat{\section}
  {\normalfont\fontsize{12}{15}\bfseries}{\thesection.}{1em}{}
\titleformat{\subsection}
  {\normalfont\fontsize{12}{15}\bfseries}{\thesubsection.}{1em}{}
\titleformat{\subsubsection}
  {\normalfont\fontsize{12}{15}\bfseries}{\itshape}{1em}{}
  
\titlespacing*{\section}
  {0pt}{1\baselineskip}{0\baselineskip}
\titlespacing*{\subsection}
  {0pt}{0\baselineskip}{0\baselineskip}
\titlespacing*{\subsubsection}
  {0pt}{0\baselineskip}{0\baselineskip}

\usepackage{caption}
\usepackage{pdfpages}

\usepackage{multirow}

\usepackage{subcaption}

\usepackage[english]{babel}

\usepackage{picture,xcolor}

\usepackage{lipsum}

\usepackage{cleveref}

\usepackage{comment}

\usepackage[english]{babel}
\usepackage[square,numbers]{natbib}
\usepackage{array}

\addto{\captionsenglish}{%
    }
\title{\bfseries \normalsize Mobile Phone Application Data for Activity Plan Generation}
    
\author[1]{\c{C}a\u{g}lar Tozluo\u{g}lu*}
\author[1]{Yuan Liao}
\author[1]{Frances Sprei}

\affil[1]{Department of Space, Earth and Environment, Chalmers University of Technology, Gothenburg, Sweden}

\date{\vspace{-5ex}}

\usepackage{algorithmic}
\usepackage[ruled]{algorithm2e}

\begin{document}

\maketitle

\textbf{* Corresponding author.}

\vspace{-2ex}E-mail address: caglar.tozluoglu@chalmers.se (\c{C}. Tozluo\u{g}lu).

\vspace{4ex}

\section*{Abstract}

Activity-based models in transport are crucial for providing a comprehensive and realistic understanding of individuals' activity-travel patterns, improving demand forecasting, policy analysis, and land use and transportation planning.
While travel surveys have long served to develop activity-based models with complete activity-travel plans, they are often costly to collect and have small sample sizes. 
Mobile phone application data, one example of emerging mobility data sources, offers an alternative with broader geographical and population coverage over extended periods, which remains under-exploited in activity-based models. 
However, the challenges of using these data include sampling biases in the population coverage and data sparsity at the individual level due to intermittent and irregular data collection.
To synthesise activity-travel plans, we propose a novel model that combines mobile phone application data with travel survey data, addressing the limitations of each data type while leveraging their strengths. 
Our generative model simulates multiple average weekday activity schedules for over 263,000 individuals living in Sweden, approximately 2.6\% of Sweden's population. 
These schedules include activity sequences, types, start/end times, and locations, incorporating daily plan variability. 
In this model, we propose a temporal-score approach to improve the state-of-the-art home and work location identification approaches among our designs to realistically synthesise activity-travel plans. 
We assess the generative model's performance against an existing large-scale agent-based model of Sweden (SySMo) and a dummy model using only mobile application data. 
The generated activity-travel plans are comparable to the SySMo model's output and significantly surpass the dummy model's results, suggesting the proposed model's capability to generate reasonable daily activity-travel schedules. 
The proposed model is adaptable to other regions with similar travel surveys and emerging data sources, like call detail records, advancing the use of these data for activity-based models in a cost-effective, easily updated manner.

\textbf{Keywords}: Mobile phone application data; Big data; Activity-based modelling; Daily activity-travel plans.

\vspace{60ex}

\section{Introduction}

Understanding individuals' mobility patterns is crucial for effective policy planning and service provision in transportation, healthcare, and urban development. 
In recent decades, activity-based travel demand modelling has become an important part of human travel behaviour modelling studies \citep{hilgert2017modeling}. 
This approach offers a better portrayal of human behavioural foundations influencing travel decisions compared to traditional population models. 
Travel surveys and diaries are commonly used as the principal data sources for these models, providing complete activity-travel plans and rich socio-demographic data \citep{huang2019transport}. 
However, this data often comes with challenges, e.g., substantial implementation costs \citep{jiang2017activity}, under-reporting of trips \citep{bricka2006comparative}, lower response rates among certain demographic groups \citep{ogle2005georgia}, and small sample size. \par

Mobile phone application data is a relatively new big data source and particularly valuable in transportation studies due to the precise location information enabled by the GPS functionality of the smartphone \citep{lee2016evaluation, rodr2024using}.
This data is collected by capturing the geographical locations of phone users as they interact with various mobile applications. 
The data enables large-scale observations of millions of individuals at a high spatial resolution over long time periods; this capability can be challenging and costly to collect with other data collection methods. 
Despite its potential, mobile application data presents significant challenges like most emerging data sources, e.g., call detail records (CDRs). 
These challenges include sampling biases due to uneven representation of different demographics and data sparsity caused by users not interacting with their phones, making it difficult to directly extract individuals' complete activity-travel patterns.

Although a few studies \citep{alexander2015origin, jiang2017activity, grujic2022combining} have shown the benefits of similar big datasets, e.g., CDRs for exploring individuals' daily mobility patterns, the use of mobile phone application data in activity-based models is under-explored. 
Moreover, these studies often overlook significant challenges, such as data sparsity, which limits the production of representative activity-travel patterns. 
At last, a notable gap is the lack of a generative model transferable to other similar big data sources.

Our study introduces a generative model that synthesises activity-travel plans by combining mobile phone data with travel survey data.
This novel approach addresses the limitations inherent to these data sources, e.g., the sparsity of emerging big data and the small sample size of traditional travel surveys, while leveraging their strengths, e.g., extensive observation periods of big data and complete one-day activity-travel plans of surveys. 
The proposed model simulates multiple average weekday activity schedules for over 263,000 individuals, approximately 2.6\% of the population in Sweden. 
These plans are characterized by activity sequences, types, start/end times, and locations, incorporating daily variability in specific plan attributes. 
We also propose a temporal-score approach to improve the state-of-the-art home and work location identification approaches by considering the variations in activity engagement at different observation times.  \par

To assess the generative model's performance, we compare it against the SySMo model \citep{tozluouglu2023synthetic}, a large-scale agent-based model and a dummy model using only mobile application data (see details in \ref{sec:dummy}). 
While the generative models' results are comparable to the SySMo model, they significantly surpass the dummy model's results. 
These comparisons highlight our approach's capability to generate realistic activity-travel schedules. 
The code for the generative models is available at \url{https://github.com/tozlucaglar/MAD4AG}, and the methodology is adaptable to other geographical contexts by leveraging similar big data and survey data.

The remainder of this paper is structured as follows: Section \ref{sec: literature} reviews the related literature. 
Section \ref{sec: methodology} describes the generative model for synthesizing individuals' activity-travel plans. 
We introduce the model evaluation methods in Section \ref{sec: evaluation}. 
Section \ref{sec: results} presents the generative model's results and compares them with other models. 
In Section \ref{sec: discussion}, we summarise our key findings, discuss the limitations of the generative model and provide suggestions for future work.

\section{Related work}\label{sec: literature}

The rise of big data sources, e.g., call detail records (CDRs), location-based social networks (LBSNs), smart-card transactions, and mobile phone application data, offers a promising alternative to develop activity-based models, traditionally reliant on travel surveys. 
These data provide a vast population and geographic coverage with superior spatial and temporal resolution \citep{lee2016evaluation}. 
However, these emerging sources often provide only partial whereabouts instead of detailed movement trajectories and do not represent the general population. 
Therefore, synthesising activity plans from big data usually faces significant challenges: 1) \textbf{incomplete records of individuals' activity travel schedules}, and 2) \textbf{population biases}.

Several studies directly use mobile phone geolocation data to generate individuals’ daily activity-travel schedules but often overlook data sparsity at the individual level. 
For example, \citet{jiang2017activity} identified individuals' activity schedules from CDR data without modelling the temporal patterns of activity participation, leaving sparsity in CDR records untreated. 
Similarly, \citet{bassolas2019mobile} and \citet{grujic2022combining} introduced pipelines to generate individuals' activity-travel patterns from mobile data without properly debiasing the data.
These studies struggle to capture full-day activity patterns realistically due to insufficient attention to data sparsity. 
This sparsity primarily arises because data is recorded only when individuals interact with their phones. 
Such intermittent and passive data collection leads to periods of uncertainty about users' whereabouts when they are not using their devices \citep{ficek2012inter, hoteit2014estimating}. 
Consequently, this sparsity complicates the direct utilization of big data without additional data sources.

To address the sparsity issue in emerging data sources, \citet{alexander2015origin} employed a travel survey as an auxiliary data source and identified individuals' daily schedules from CDR data. 
The study inferred individuals' activity locations and, between the locations, defined trips categorised into home-based work, home-based other, and non-home-based. 
To overcome the sparsity and capture the realistic temporal pattern of trips, they randomly sampled the trips' departure times based on the travel survey trip departure time distribution. 
However, this approach does not model participation in daily activities. 
It directly extracts activity plans from incomplete mobile phone data, leading to less realistic temporal patterns of activities.

Primary activities, e.g. home, work, and school, influence individuals' daily plans and are crucial for developing activity-based models \citep{arentze2000albatross, Bowman2001Activity-basedSchedules}.
Despite mobile phone data's spatial and temporal sparsity, its large volume and extended observation period make it feasible to infer certain components of the activity schedules, e.g., primary activities' locations. 
Existing studies mostly rely on temporal rules to identify locations, assuming individuals will be in specific places during predefined hours. 
For instance, \citet{chen2014traces} identify individuals' home and work activity locations by analyzing daytime and nighttime periods, achieving success rates of 70\% and 65\% for home and workplace inference within 100 m compared with ground truth data.
\citet{sadeghinasr2019estimating} detect home locations during the nighttime from 8:00 pm to 4:59 am.
Additionally, \citet{ahas2010using}, \citet{kung2014exploring}, and \citet{yu2020mobile} infer individuals' home and work locations by applying distinct time frames using CDR data, providing essential information for evaluating commuting dynamics. 
While these assumptions are often valid, exceptions exist, and individuals' patterns of being at home or work can vary.

Among the emerging geolocation data sources on human mobility, mobile phone application data is relatively new and has mostly been used in population-based modelling. 
This data shares characteristics similar to CDR data, e.g., collection methods through individuals' interactions with their mobile phones and data sparsity at the individual level due to intermittent data collection.
Researchers have utilized mobile phone application data to generate OD matrices, illustrating trip distributions between locations \citep{zhang2019mobile, yu2020mobile_bike}.
\citet{zhang2020mobile} and \citet{yu2020mobile_bus} used trajectories from mobile phone GPS datasets to identify individuals’ travel modes. 
\citet{woodard2017predicting} predicted travel time variability at the trip and the road network link levels.
Despite these studies,  the application of mobile phone application data in activity-based models remains under-explored. 
Given its spatial precision compared to other big data sets, mobile phone data has the potential to accurately deduce activity locations, thereby enhancing the realism of activity-based models.

The existing literature highlights the benefits of emerging data sources in transportation studies. 
Despite the progress in leveraging these data, considerable gaps remain in generating activity-travel plans. 
A significant challenge lies in addressing the sparsity of big data. 
Moreover, integrating mobile phone application data into activity-based models holds potential due to its spatial precision, large population coverage, and cost-effectiveness, while it remains under-explored. 
Additionally, there is a need for more rigorous methods to identify activity locations, moving beyond evaluations based on temporal rules and visit frequencies. 
These enhancements would improve activity-based models' realism and advance our understanding of human travel behaviour.

\section{Model design}\label{sec: methodology}

We propose a generative model that simulates multiple average weekday activity-travel patterns by combining massive mobile phone GPS records with the national travel survey data (detailed in Section \ref{seca:data} Data). 
The model generates individuals who perform activities to fulfill various personal and social needs.
The generated activities inform the overall travel demand of the population \citep{kitamura1984model}, and can be applicable in simulations for infrastructure planning, public transportation optimization, emergency response planning, environmental impact assessments, and economic analysis of transport policies. \par

\begin{figure}[htbp!]
    \centering
    \includegraphics[width=1\textwidth]{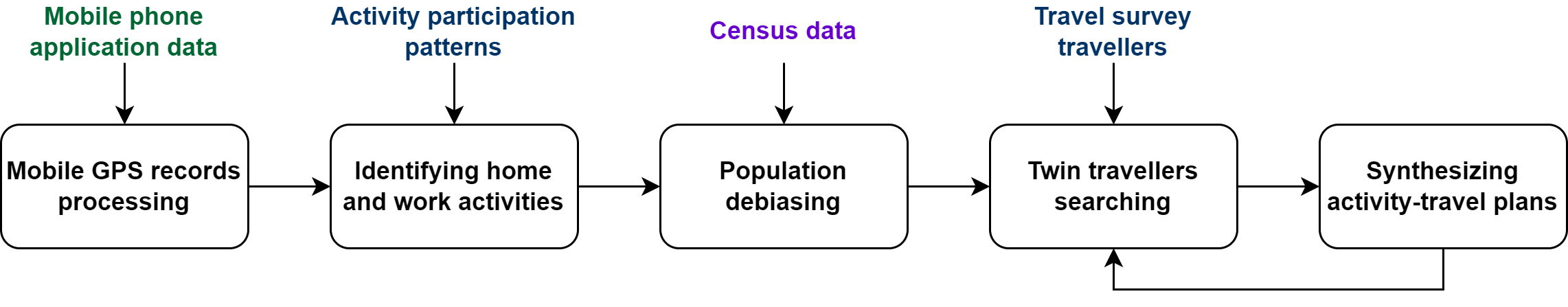}
    \caption{Generative model for synthesising activity-travel plans. \textmd{Travel survey inputs are marked with blue. Census data refer to the statistics related to Demographic statistics areas (DeSO zones), specified in Section \ref{data:deso}.} }
    \label{fig: flowchart}
\end{figure}

The proposed generative model includes five modules (\Cref{fig: flowchart}).
First, we process mobile phone GPS records for further use (Section \ref{sec:data_prep} Mobile phone data processing). 
Second, we identify the locations of individuals' primary activities, i.e., home and work. 
We assume each individual has at least one home activity on an average day.
Third, we design individual weights to have a representative sample of the Swedish population.
Fourth, we start searching for twin travellers in the travel survey by calculating the matching probabilities between people in the mobile phone application data and the travel survey participants. 
The matching probability is defined based on their demographic, activity patterns, and trip attributes.
For each individual in the mobile phone dataset, we sample one twin traveller from the identified travel survey candidates. 
Finally, we generate activity-travel schedules for average weekdays using the selected twin travellers' daily activity sequence and individuals' activity locations in the mobile phone dataset. 
The proposed model is capable of synthesizing different activity-travel plans for each agent, producing daily variations in individuals' activity-travel patterns.
By iteratively finding twin travellers and activity locations, the model generates multiple average weekday simulations for each individual in the mobile phone dataset.

\subsection{Mobile phone data processing}
\label{sec:data_prep}

Through the GPS records from 322.919 mobile phone devices in Sweden, we first detect stops, which are defined as the geolocations where individual devices have spent more than 15 minutes.
Based on the detected stops, we further identify activity records and their clusters. 
To improve the data quality for synthesising activity-travel plans, we also implement a few data filters, e.g.,  individuals with very few or outside Sweden records, etc.

\subsubsection*{Stop detection}

To generate synthetic activity plans, it's essential to identify when and where individuals are engaged in various activities. 
This process begins by detecting "stops" from individual geolocation trajectories, which consist of user ID, latitude, longitude, and time information. 
A stop is where an individual remains for at least 15 minutes, characterised by latitude, longitude, start, and end times. 
These stop observations form the basis for creating activity clusters, identifying primary activity locations, and synthesising activity-travel plans in the proposed model. \par

Various algorithms can detect stops, as outlined in the literature \citep{ester1996density, aslak2020infostop}. 
For this study, we've chosen the Infostop algorithm \citep{aslak2020infostop}, known for its robustness against measurement noise, scalability with large datasets, and ability to process multi-user data concurrently. 
Details about the chosen algorithm and its parameter selection are discussed in Section \ref{seca:stop_detection} Stop Detection \citep{liao2024uneven}. 
The geolocations where individuals spent more than 15 minutes were observed over seven months in 2019 in Sweden. \par

\subsubsection*{Identifying activity records}

\textbf{Creating activity clusters.} Activity space is widely used to describe the geographical area where individuals conduct their daily activities within a time frame \citep{golledge1997spatial, schonfelder2003activity, miller2004activities}. 
It encompasses the locations and spaces where people live, work, study, socialise, and engage in various other activities.
After pinpointing the geolocations where individuals spend significant time, we form activity clusters to represent their daily activity spaces. 
This approach facilitates the creation of more representative daily activity-travel plans with less noise.\par

The geolocation data from mobile applications reveal distinct activity-travel patterns for individuals, e.g., one area with many activity locations in an urban area and one around a remote summer place. 
Contrary to earlier studies suggesting scale-free human movements \citep{brockmann2006scaling, gonzalez2008understanding, song2010modelling}, recent research shows that human mobility has definable scales, with individuals moving within and between different spaces confined by these scales \citep{alessandretti2020scales}. 
Therefore, by examining individuals' activities and travel patterns, we identify activity clusters that reflect their daily mobility scales, excluding rare events such as residential mobility from one area to another. \par

To determine individuals' activity clusters, we apply the Density-Based Spatial Clustering of Applications with Noise (DBSCAN) algorithm, which has been widely used in the literature \citep{sadilek2012modeling, stalla2016anonymous, wojtusiak2021location}.
The DBSCAN algorithm clusters points based on proximity, requiring a minimum number of points ($MinPts$) within a specified radius ($\varepsilon$) \citep{ester1996density}. 
We set the $\varepsilon$ parameter to 200 km based on the 99th percentile of travel distances from the travel survey and the $MinPts$ parameter to 2 to ensure the detection of at least one additional activity alongside the home activity within each cluster.
This process allows us to establish spatial scales where we infer typical weekday activity patterns, treating activity clusters that are distant from each other separately.
For example, an individual residing in one location for three months and then another for three months would have two activity spaces, which can be identified and analyzed separately by the model.

\textbf{Building-level activity records.} After identifying activity clusters, we group stops of each device in each activity cluster to determine where individuals engage in daily activities, e.g., home, work, and others, thereby generating activity records.
Each stop offers a glimpse or segment of individuals' activities at a given location.
To identify activity records, the DBSCAN algorithm is applied to the stops' GPS coordinates, with the $\varepsilon$ parameter set to 100 m, reflecting the spatial resolution observed in the data.
This method groups the stops based on proximity.\par

However, one activity may end up in different locations for two reasons: noise from GPS records and movement within the activity location.
First, factors such as GPS receiver type, signal quality, and environmental conditions cause variations in spatial resolution of stops, ranging from centimetres to hundreds of meters \citep{thierry2024comparing}. 
Second, the small movements of individuals while engaged in activities further disperse the stop locations, e.g., moving between offices at work. 

To address the scattering of identified activity records around the buildings, we generate building-level activity records that link the activity records with buildings for use in further analysis.
We derive representative GPS coordinates for activity locations by calculating the mean values of the coordinates of the stops within each group. 
Subsequently, these calculated coordinates are snapped with the nearest building to represent the recorded activities' geolocations.

\subsubsection*{Filtering data}
We select individuals and their activity locations for further analysis based on specific criteria to guarantee the data quality.
Activities that exceed 12 hours are excluded, as are those outside of Sweden.  
We also exclude individuals with fewer than seven active days, where an active day is defined as having at least one recorded activity.
Activities collected on weekends and holidays are not considered.
Lastly, individuals with fewer than two unique activity locations within their activity clusters are removed from the analysis. 
These criteria ensure a focused dataset for our study. \par

After processing, the data contains 287,217 individuals with 12,740,559 activity records distributed across 312,901 identified activity clusters. 
Table \ref{tab:analysis} shows statistics about individuals and their activity records.

\begin{table}[hb!]
\small
\center
\caption{Descriptive statistics of activity records. An active day is a day with at least one recorded activity.}
\label{tab:analysis}
\begin{tabular}{llllll}
 \\ \hline
Attribute                  & min & 25\% & 50\% & 75\% & max  \\ \hline
No. of activities               & 2   & 10   & 21   & 48   & 1129 \\
No. of unique activity locations    & 2   & 3    & 4    & 10   & 374 \\
Median activity duration (min) & 16  & 180  & 182  & 192  & 717  \\ \hline
\end{tabular}
\end{table}

\subsection{Temporal-score approach for inferring primary activities}\label{sec:primary}

We infer primary activity locations for individuals in the mobile phone application data. 
Primary activities (or anchor activities) are critical activities, e.g., work and school, whose locations and timings are determined independently of other activities \citep{arentze2000albatross, Bowman2001Activity-basedSchedules}. 
These activities play a role for individuals to plan their secondary activities, including shopping, leisure, and social visits. 
In this study, we define two categories under primary activities: home and work activities.
The work activity category also includes educational activities. 
Existing methods often rely on temporal rules to identify residential locations \citep{yabe2022mobile, vanhoof2020performance, jiang2017activity}. 
These rules include that individuals are mostly at home during nighttime and that the most frequently visited place is the individual's home. 
Although these assumptions hold in many cases, some exceptions exist, and the patterns of being home and working may vary among individuals. \par

To infer \textbf{home locations}, we propose a temporal score method that accounts for the varying likelihood of presence at an activity location across different hours of the day. 
We infer the primary activities based on the activity cluster where they are situated.
We first calculate the number of visits to each activity location during nighttime over the observation period. 
The definition of nighttime varies across studies, e.g., 8:00 pm to 4:59 am \citep{sadeghinasr2019estimating} and 7:00 pm to 7:59 am \citep{vanhoof2020performance}. 
In this study, we set a temporal rule for nighttime based on the travel survey data, defining it as the hours in a day when the rate of home activity participation surpasses 0.8 of the population, i.e., the hours where 80\% of the travel survey participants are home. 
Accordingly, the nighttime starts at 6:00 pm and ends at 7:59 am the following morning. 
Each visit to an activity location during this time frame is counted as one per day, regardless of whether the location is visited one or more times. 
The activity locations being visited fewer than three times during the data-informed nighttime are excluded from further analysis. \par

Next, we deal with the remaining activity locations being visited at least three times during the pre-defined nighttime.
In this step, we calculate a score, considering the variations at the observation time rather than equally treating the hours within the nighttime. 
For example, according to the travel survey data, more individuals are home at 3:00 am than at 7:00 am.
Therefore, an activity location visited at 3:00 am is more likely to be home than at 7:00 am.
To calculate such scores, we design weights for each nighttime hour between 6:00 pm and 7:59 am, derived from the ratio of hourly home activity participation to overall activity participation in the travel survey (Figure \ref{fig: weights}).
Accordingly, we weigh the duration of an individual's stay at each activity location during the night hours and calculate its score (detailed in Section \ref{seca:home-work_detection}). \par

We identify home activity locations with the highest score above a score threshold and being at least visited three times during the data-informed nighttime.
The score threshold is set to 10 by evaluating the inferred residential location's building type. 
The threshold is chosen to balance a reasonable share of home activity locations at residential buildings and sufficient data. 
We identify the residential locations of over 263,000 individuals in the mobile phone dataset.
Individuals whose home locations could not be inferred are excluded from further analysis based on our assumption that every individual participates in a home activity. \par

To identify \textbf{work locations}, we adopt a similar approach to the residential location inference. 
After excluding the inferred home locations from the activity locations, we evaluate typical working hours, specifically from 8:00 am to 5:59 pm, according to the travel survey. 
We then calculate a score for each remaining activity location (detailed in Section \ref{seca:home-work_detection}). 
To ensure the quality of workplace location inference, we remove activity locations with a score below 30.
We keep the activity location with the highest score among all evaluated locations as a workplace. 
We identify the workplaces of approximately 67,000 people, accounting for 26\% of the population whose home locations are identified. \par

\subsection{Population debiasing}

To ensure a representative sample of the Swedish population, we compute a weight for each individual in the mobile phone data. 
These weights address the inherent sampling bias arising from the spatial distribution of the individuals in the mobile phone dataset across Sweden. 
To calculate weights, we first associate individuals in the mobile phone dataset with Demographic statistics areas (DeSO zones, specified in Section \ref{data:deso}), using their inferred home locations. 
We then apply these zones' population and employee sizes in the following weight design. \par

The weight calculation has two stages: First, we apply inverse probability weighting (IPW) to assign initial weights based on the ratio of the calculated population size to the adult population size in the DeSO zones. 
Second, we adjust these weights to match the distribution of employees across different zones.
We use the Iterative Proportional Fitting (IPF) technique that iteratively adjusts the elements of a matrix to ensure that the sums across each row and column align with predetermined totals \citep{deming1940least}. 
In each DeSO zone, we scale the initial weights of employees by the actual employee size in the statistics while maintaining the overall population size (Equation \ref{eqn:w_p}).
For example, if employees in a DeSO zone are expected to constitute 50\% of the population, but only 40\% are observed in our dataset, IPF addresses this discrepancy by increasing the weights of employees and reducing the weights of non-employees with a multiplier reflecting the accurate employment rate, while maintaining the population size in each DeSO.\par

\begin{align}
     &\quad W_{p,e}' \leftarrow \left(\frac{N(z, e)}{\sum_{p \in z, e} W_p}\right) \cdot W_{p,e} \notag \\
    \text{$ \forall (p \in z):$} &\quad W_p' \leftarrow \left(\frac{N(z)}{\sum_{p \in z} W_p'}\right) \cdot W_p'
    \label{eqn:w_p}
\end{align}

where $p$ denotes individuals and $W_{p}$ their initial weights, $e$ is employees. 
$N(z, e)$ represents the desired number of employees in zone $z_d$, $\sum_{p \in z, e} W_p$ is the sum of initial weights for employees in that zone, and $\sum_{p \in z} W_p $ is the sum of weights in the zone. 
$ W_p'$ reflects the adjusted value after the IPF process. \par

To prevent extreme weights, we apply the weight-trimming technique, where any weight above the threshold ($W_0$) is set to $W_0$ \citep{van2014weight}. Equation \ref{eqn:w_o} calculates this threshold.

\begin{align}
    W_0 = 3.5 \times \sqrt{1 + \text{CV}^2(W_p) \times \text{Med}(W_p)}
    \label{eqn:w_o}
\end{align}

where $W_{p}$ denotes the individuals' weights, CV is the coefficient of variance, and Med is the median value of the weights.

\subsection{Searching twin travellers}\label{sec:search_twins}

The model synthesises activity-travel plans from abundant but incomplete activity records from mobile phone data.
To do so, we first synthesise activity sequences for individuals in the mobile data, considering their observed mobility patterns and their similarity to the travel survey participants who have complete activity records.
We adopt a statistical matching approach using the travel survey data. 
Statistical matching with some variations is often applied in activity-based modelling to define daily activity patterns for synthesised individuals based on their similarity to survey participants \citep{horl2021synthetic, namazi2017unconstrained}. \par

The statistical matching approach connects each individual in the mobile phone data with one travel survey participant, defined as his/her "twin traveller." 
We first group individuals in the mobile phone data and the travel survey data based on key mobility attributes, e.g., employment status and average trip distance, and then calculate a matching likelihood probability between them. 
Using these probability results, we pair each individual in the mobile phone data with one twin traveller in the travel survey participants considering their shared mobility attributes. \par

\subsubsection*{Categorization of travellers}\label{sec:cate_trav}

We group individuals in the mobile phone and travel survey data based on their demographic, activity, and trip attributes.
The considered attributes are the individual's residential region (Svealand, Götaland, or Norrland), urban density of residence (high or low), employment status (yes or no), average trip distance between home and activities (long or short), and commuting distance (long or short). 
For the individuals in the mobile phone data, we use all the attributes. 
For the travel survey participants, however, we adopt a step-wise grouping approach due to the limited sample size. 
The detailed definitions of these variable levels are in Section \ref{seca:search_twins} Searching twin travellers. \par

After this step, all individuals in the mobile dataset are assigned a group ($c_i$ for individual $i$). 
For example, individuals with a high average trip and commuting distance residing in a low urban density area in Götaland. \par

\subsubsection*{Sampling twin travellers}

Classifying mobile phone users and travel survey participants into the same category of travellers creates a common basis for the next step: identifying a twin traveller for each individual in the mobile phone dataset within each category.
Instead of randomly selecting the twin traveller within the same category, we calculate matching likelihood probabilities for each travel survey participant, considering the resulting distributions of commuting and activity patterns at the aggregated level. \par

We calculate the matching likelihood probabilities using the IPF technique.
We start by assigning a uniform probability to each individual within a category. 
This initial probability is inversely proportional to the total number of individuals in that category. 
We then adjust these probabilities using multipliers that reflect the distributions of commuting and activity sequences observed in the survey data.
We iteratively scale the probabilities to match the cumulative probability distribution with the respective marginal distributions. 
After each adjustment, we also normalise the probabilities for each individual to ensure that the sum of probabilities within each category equals the total number of individuals in that category. 
This iterative fitting process continues until the adjusted probabilities converge, indicated by negligible changes between subsequent steps. 
Equations \ref{eq:group_maintain1}-\ref{eq:group_maintain3} show the calculation of the matching likelihood probabilities.\par

\begin{align}
 p^{(t+1/3)}_{i,k,s} = p^{(t)}_{i,k,s}\frac{D_k}{\sum_{i,s} p^{(t)}_{i,k,s}} \label{eq:group_maintain1} \\
 p^{(t+2/3)}_{i,k,s} = p^{(t+1/3)}_{i,k,s}\frac{S_s}{\sum_{i,k} p^{(t+1/3)}_{i,k,s}} \label{eq:group_maintain2} \\
p^{(t+1)}_{i,k,s} = \frac{p^{(t+2/3)}_{i,k,s}}{\sum_{i,k,s} p^{(t+2/3)}_{i,k,s}} \label{eq:group_maintain3} 
\end{align}

where $p_{i,k,s} $ is the probability assigned to each individual $i$, commuting type $k$, and activity sequence type $s$ in category $c_i $, $D$ is the vector of marginal probabilities for commuting, and $S$ marginal probabilities for activity sequences. After initializing the individual weight $p^{(0)}_{i,k,s}$ for individual $i$, we iteratively adjust the weight $p_{i,k,s}$ to fit the marginal probabilities $D$ and $S$. In each iteration $t$, we perform the row adjustment for commuting probabilities (Equation \ref{eq:group_maintain1}), the column adjustment for sequence probabilities (Equation \ref{eq:group_maintain2}), and normalisation (Equation \ref{eq:group_maintain3}). This iterative fitting process continues until the adjusted probabilities converge, indicated by negligible changes between subsequent steps. \par

To find twin travellers for individuals in the mobile data, we first identify candidate individuals within each group for the travel survey participants. 
Using the calculated matching probabilities, we randomly sample a set of survey participants to equal the number of individuals in the mobile dataset for each group. 
Secondly, we rank individuals from both groups in ascending order by their average trip distance and match them one-to-one. 
During the matching process, we ensure that individuals without identified secondary activity locations are not matched with survey participants who have such activities in their sequence.  
Ultimately, we assign the individuals in the mobile phone dataset with the activity types, start and end times, and sequences of their twin travellers in the travel survey. \par

\subsection{Synthesizing activity-travel plans}\label{seca:syn_plans}

After identifying a twin traveller for each individual in the mobile phone dataset, we have activity schedules assigned.
In this section, we show how we continue assigning primary and secondary activity locations to the corresponding activity schedules.

\subsubsection*{Location assignment of other activities}

Other activities (secondary activities) are all activities apart from primary activities in a daily activity schedule, e.g., shopping, leisure, etc. 
The mobile phone dataset, covering seven months of observation, features abundant other activity locations for individuals, providing valuable variations in travel demand induced by secondary activities. 
On average, each individual has 6.6 other activities, with approximately 18\% individuals having more than ten other activity locations. 
We thus need to determine which other activity locations are visited on an average weekday in the synthesised activity-travel plans. \par

For all other activities in the individual's schedule, we select the corresponding locations by considering their distances to their home location and the number of other activities in the individual's activity sequence. 
We define the probability of selecting one other location as below:

\begin{align}
    P_o\propto\frac{f_o}{\log(d_{o,h} + 1)}
    \label{eqn:coef}
\end{align}

where $P_o$ represents the probability of selecting one other activity location, $f_o$ the frequency of visits to this location, $h$ the individual's home location, and $d_{o,h}$ the distance between them. \par

The individuals' probability of visiting each other activity location is normalised by all other location records so that the probability of visiting all adds up to 1. 
Using the calculated probabilities, we randomly sample other activity locations for all other activities in the individual's activity sequence.

\subsubsection*{Completing activity-travel plans}

Now, we assign an activity sequence for each individual in the mobile phone dataset, including activity types, start/end times, and where these other activities occur, if any. 
We have also previously inferred the primary activity locations, i.e., home and work, from the steps in Section \ref{sec:primary}.
We first place these primary activity locations into the individuals' activity schedules. 
We then allocate the selected other activity locations to the corresponding activities in the schedule. 
The other activities are arranged chronologically and positioned based on proximity to the most recently placed primary activity.
In other words, the other activity location is assigned if it is the closest to the preceding primary activity location. \par

The proposed generative model is capable of generating multiple average weekday activity-travel plans for one individual in the mobile phone dataset. 
One can repeat the steps of searching for twin travellers (Section \ref{sec:search_twins}) and synthesising activity-travel plans (Section \ref{seca:syn_plans}) to get a variety of plans for individuals. 
At the individual level, variations occur across all elements of activity schedules, i.e., activity sequences, type, start/end times, and other activity locations, while maintaining the residential and workplace locations. 
Despite these variations at the individual level, the activity patterns at the population level remain robust by design.

\section{Model evaluation methods}\label{sec: evaluation}

We assess the performance of the generative model by comparing its activity-travel schedules with the travel survey data and other similar activity-based models.
The performance is evaluated by reflecting the key aspects of travel demand, e.g., daily activity sequence distribution, temporal activity engagement patterns, and trip distance distribution.
We include comparisons with a dummy model, which adopts a simplified methodology using only mobile phone datasets to generate average weekday activity-travel schedules, and the Synthetic Sweden Mobility (SySMo) model, a large-scale agent-based transportation model of the Swedish population.
These comparisons provide insights into the generative model's ability to integrate mobile data with survey data to generate realistic simulations.

\subsection{Dummy model}\label{sec:dummy}

To serve as a baseline for comparisons with the generative model, we present a dummy model that adopts a simpler methodology and uses only the mobile phone dataset. 
The model generates average weekday activity-travel schedules characterised by activities' type, start and end times, duration, sequence, and location. \par

The model includes three steps: mobile phone data processing, inferring activity locations, and generating activity schedules. 
In the mobile phone data processing step, we adopt the approach used in our generative model, detecting stop points where the individual remains stationary for significant periods and identifying activity records from these stops.
In the second step, the model infers primary and secondary activity locations categorised using the state-of-the-art approaches instead of the temporal-score approach in this study (Section \ref{sec:primary}). 
The primary activities are home and work, and the secondary activities are all remaining activities, e.g., shopping and restaurant visits. 
The model determines primary activity locations by counting the number of visits to each location during designated hours: nighttime for homes (6 pm to 7:59 am) and off-nighttime hours (8:00 am to 5:59 pm) for workplaces. 
The location with the highest visit count during these hours is assigned as the respective primary activity location. 
The remaining abundant activity locations visited during the observation period are categorised as secondary activity locations.\par

The final step includes generating individual activity-travel schedules from the mobile phone dataset. 
For each individual, the model calculates the hourly frequency of visits to each activity location over the observation period. 
The model assigns the most frequented location during each specific hour of the day as the activity for that time. 
Using previously inferred activity types, the model constructs daily activity schedules. \par

The dummy model relies solely on mobile phone geolocation data and generates activity-travel schedules without considering the sparsity of the dataset as well as the population biases.
It provides a data-driven representation of the spatio-temporal mobility patterns of its covered mobile devices.
The methodological differences between the proposed generative model and this dummy model are summarized in Table \ref{tab:model_met_comparison}.

\begin{table}[h!]
\centering
\caption{Model methodology comparison between dummy and proposed generative model}
\begin{tabular}{|p{4cm}|p{5cm}|p{5cm}|}
\hline
\textbf{Model Component} & \textbf{Dummy Model} & \textbf{Proposed Model} \\ \hline

Primary activity \newline inference & Infers primary activities based on visitation frequency. & Considers stay durations using the temporal-score approach informed by survey data. \\ \hline

Secondary activity \newline inference & Uses only secondary activity locations with the highest hourly visitation frequencies. & Incorporates all observed secondary activity locations in the data, considering their visitation frequency and distance from primary activities. \\ \hline

Activity sequence \newline generation & Derives activity sequence distribution from the input data. & Mirrors activity sequence distribution in the travel survey data. \\ \hline
Activity schedule \newline generation & Generates activity sequences and start/end times based on hourly visitation frequency, relying on abundant but incomplete activity records. & Copies activity sequences and start/end times from twin travellers with complete activity records. \\ \hline

Simulation & Generates a single average day activity-travel patterns. & Generates multiple average day activity-travel patterns using its probabilistic methodology. \\ \hline
\end{tabular}

\label{tab:model_met_comparison}
\end{table}

\subsection{Synthetic Sweden Mobility (SySMo) model }\label{sec:sysmo}

The Synthetic Sweden Mobility (SySMo) model \citep{tozluouglu2023synthetic} is a large-scale agent-based transportation model. 
It simulates the Swedish population and their transport behaviours on an average weekday.
The model generates a synthetic replica of over 10 million individuals with socio-demographic attributes, e.g., age, gender, civil status, etc.
These attributes are the basis for travel demand generation.
Using an activity-based approach for demand generation, the model assigns daily activity-travel schedules, including the type of activity, start-end time, duration, sequence, the location of each activity, and the travel mode between activities to each individual in the population. \par

The model's results are evaluated against official statistics from Statistics Sweden and Trafikanalys and are publicly available in a data repository \citep{SySMo_repo}. The model outputs, which include agents' socio-demographic information and mobility patterns, have served as a base for other research, e.g., \citet{liao2023impacts}, \citet{tozluouglu2024potential}.

\subsection{Performance metrics}\label{sec:metrics}

To evaluate the performance of the generative model, we conduct a series of comparative analyses against the dummy and the Synthetic Sweden Mobility (SySMo) models. 
These analyses focus on three key aspects of human mobility: daily activity sequences, temporal activity participation patterns, and trip distances.
We choose these comparisons to provide a comprehensive assessment of the generative model's ability to represent the mobility behaviour of the Swedish population. \par

To quantify the similarity of distributions across different models, we use the Jensen–Shannon ($JS$) distance. 
The JS distance is a symmetrized and smoothed variation of the Kullback-Leibler ($KL$) divergence \citep{osterreicher2003new}. 
To compute the $JS$ distance, we first calculate the KL divergence for each pair of distributions. 
The $JS$ distance is then derived as the square root of the average of the $KL$ divergences between each distribution and the mean distribution, $M$, of the pair ($p$ and $q$ distributions). 
This method results in a distance range from 0 to 1, where 0 indicates identical, and 1 is a different distribution. 
We calculate the $KL$ divergence and the $JS$ distance as below:

\begin{align}
\begin{split}
   KL(p, q) &= \begin{Bmatrix*}[l]
     p \log(p/q)-p+q &\quad & p>0,q>0  \\
    q & \quad &  p = 0, q \geq 0 \\
    \infty &\quad & otherwise
   \end{Bmatrix*} \\
   \\
    JS(p, q) &= \sqrt{\frac{KL(p,M)+KL(q,M)}{2}} \\
\end{split}\label{eqn:JS_dist}
\end{align}

\section{Results}\label{sec: results}

This section presents the results of the generative model that synthesises activity-travel schedules by fusing the mobile phone application and the travel survey data. 
We also assess the model’s performance by comparing its outputs with various sources, including grid-level population statistics, the SySMo and dummy models, and travel survey data that served as one input for the generative model. \par

First, we present the primary activity inference results and compare them with the actual population size. 
Next, we examine the distribution of activity sequences and temporal activity participation across the models and the travel survey. 
Finally, we compare trip distances and commuting distances across the different models. 
This comparison helps to highlight the model's strengths and potential areas for improvement in synthesising activity-travel patterns using big geolocation data sources.

\begin{figure}[htbp!]
    \centering
    \begin{tabular}{lr}

        \begin{minipage}{0.55\textwidth}
            \centering
            \includegraphics[trim={15.4cm 0.6cm 1.30cm 2.25cm}, clip, width=.99\textwidth]{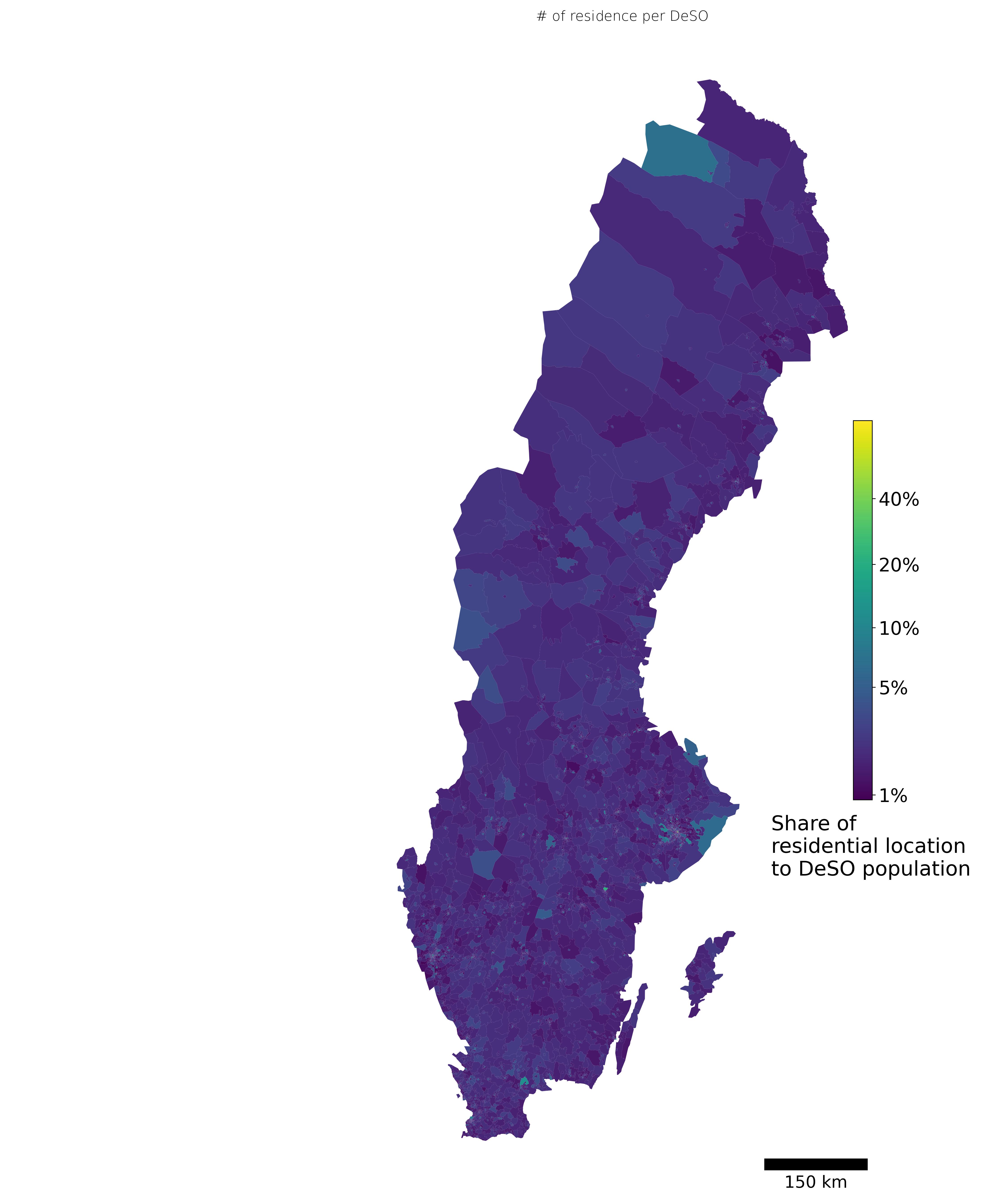} \\
            (a) 
        \end{minipage}
        &
        \begin{minipage}{0.45\textwidth}
            \centering
            \includegraphics[trim={0.6cm 0.5cm 0.5cm 0.5cm}, clip, width=0.95\textwidth]{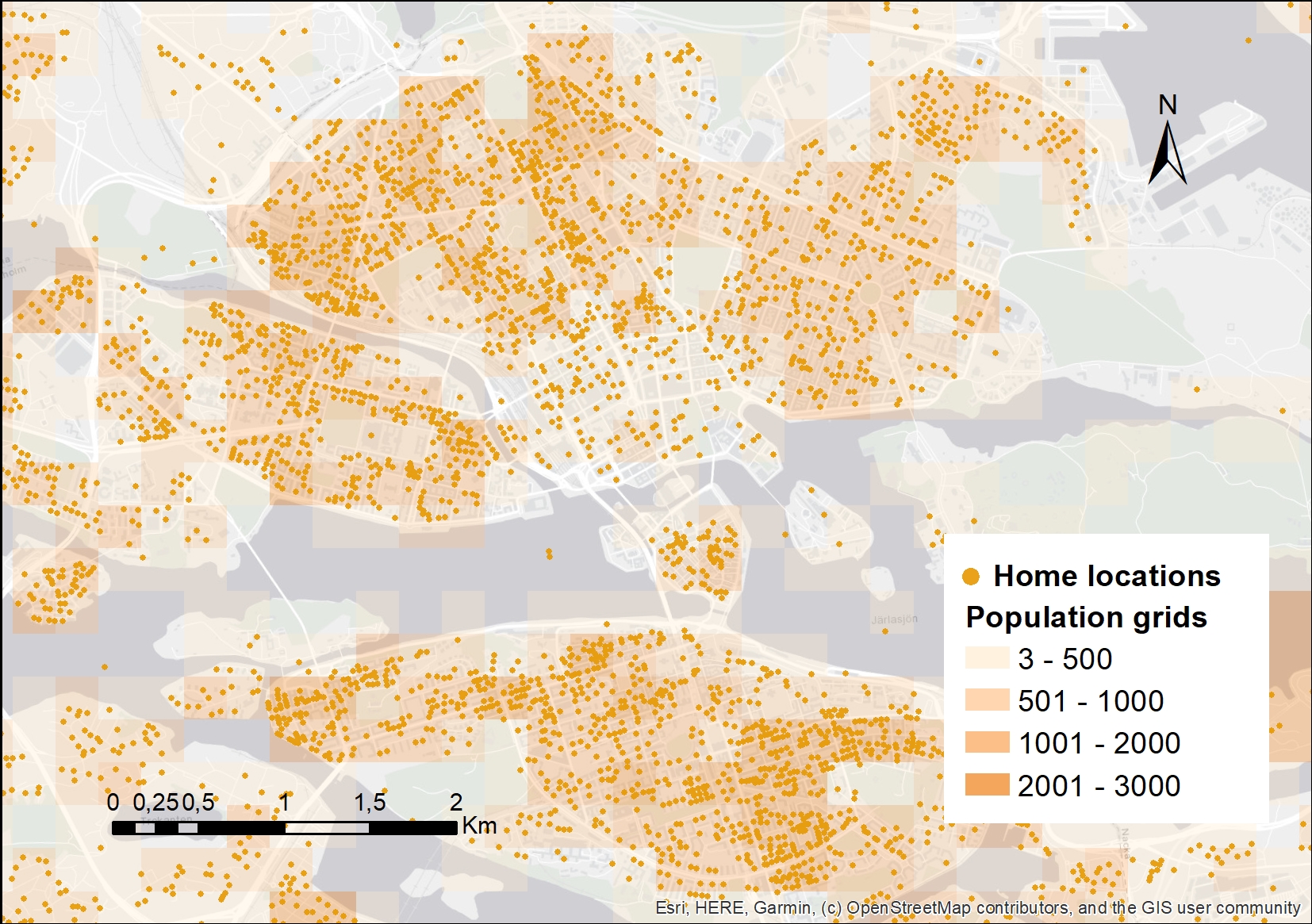} \\
            (b) \\[6pt]  
            \includegraphics[trim={0.6cm 0.5cm 0.5cm 0.5cm}, clip, width=0.95\textwidth]{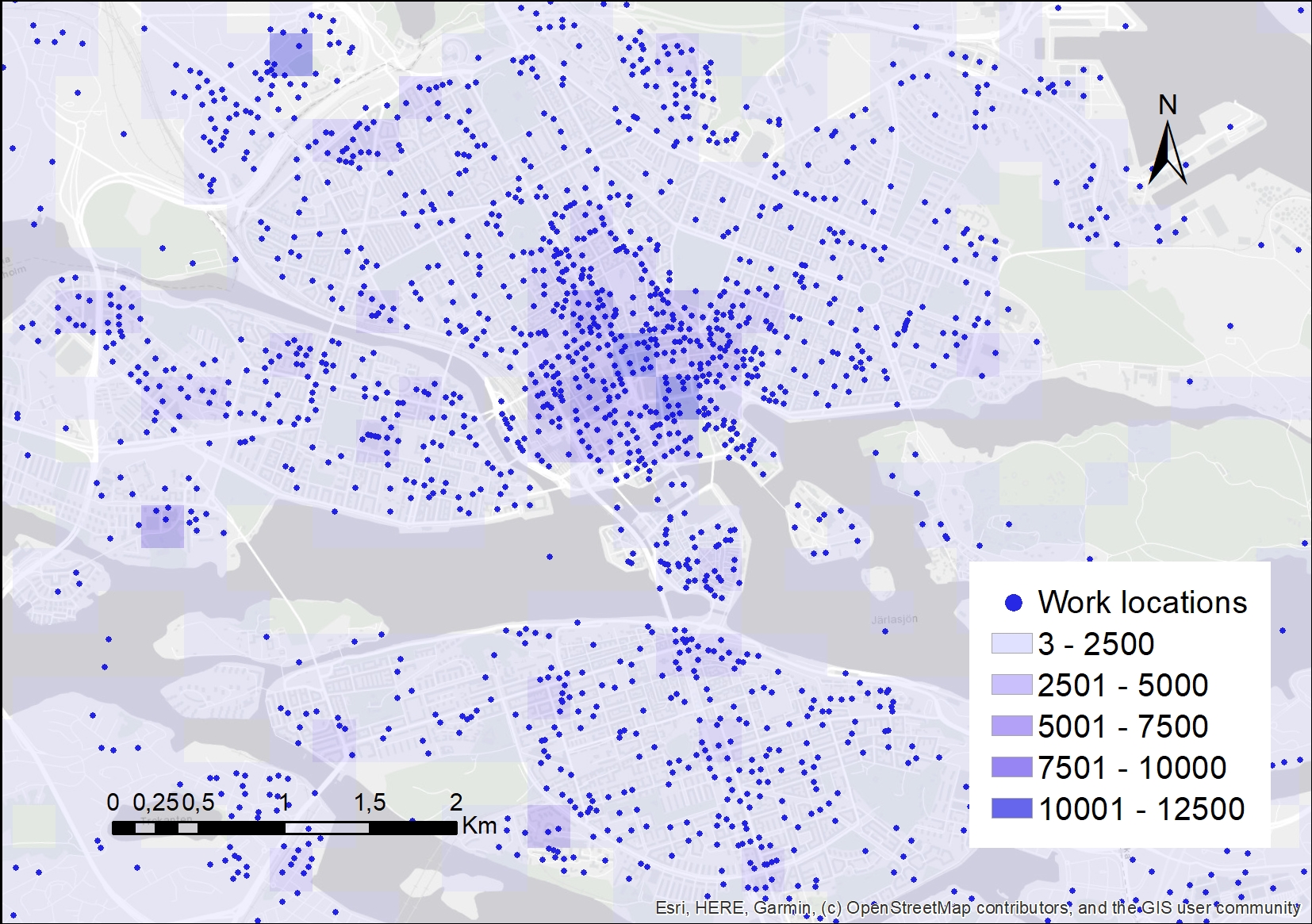} \\
            (c) 
            \\[6pt]  
            \includegraphics[trim={0.6cm 0.5cm 0.5cm 0.5cm}, clip, width=0.95\textwidth]{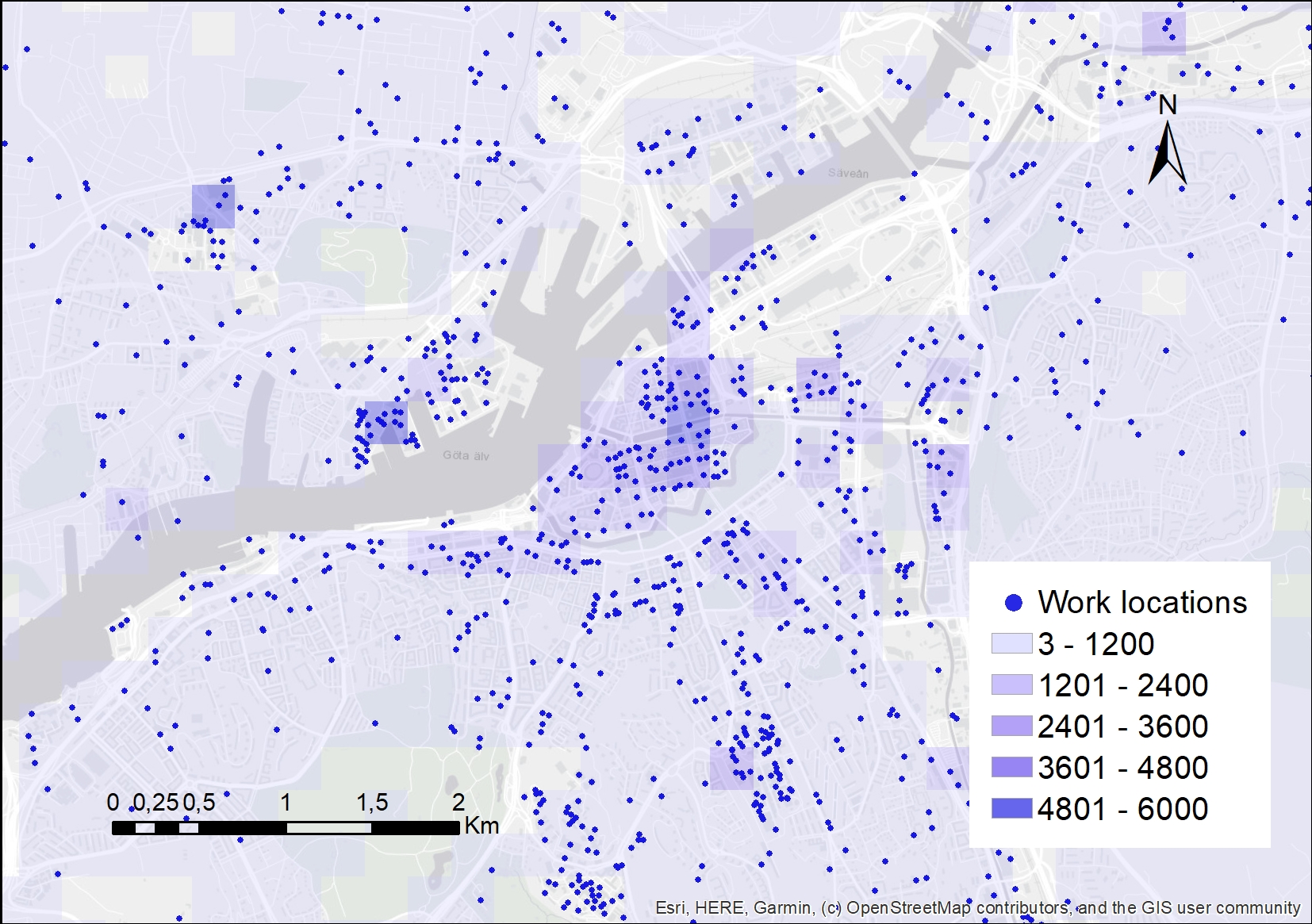} \\
            (d) 
        \end{minipage}
    \end{tabular}
    \caption{Identified home and work locations vs. census statistics: \textmd{(a) Share of the inferred home locations to the DeSO zone population (in \%). (b) Comparison of the inferred home locations with the population size in 250 m grids \citep{DeSO} across central Stockholm. (c) Comparison of the inferred workplace locations with the number of employees in 250 m grids \citep{DeSO} across central Stockholm. (d) Comparison of the inferred workplace locations with the number of employees in 250 m grids \citep{DeSO} across central Gothenburg.}}
    \label{fig:primary_activities}
\end{figure}

\subsection{Primary activity locations}

The generative model identifies the individuals' primary activity locations, i.e., home and work locations, from the mobile phone application data. 
These activities are crucial for constructing activity schedules, and their spatial distribution reflects the demographic representativeness of the data. \par

We analyse the spatial distribution of primary activities to give a more comprehensive perspective on primary activity inference.
Figures \ref{fig:primary_activities}a and \ref{fig:primary_activities}b illustrate the distribution of identified home locations compared to the actual population sizes. 
The generative model identifies home locations in each DeSO zone. 
Figure \ref{fig:primary_activities}a compares the identified home locations with the population size at the DeSO zones. 
In most zones, the share of the number of home locations in the total population is between 1\% and 5\%. 
We also calculate a Spearman correlation coefficient of 0.49 ($p$ < 0.001) between the number of identified homes and population size. 
This value indicates a moderate representation of the population's magnitude in each DeSO zone. 
Figure \ref{fig:primary_activities}b further explores the home location distribution with population density at a 250m grid level in central Stockholm, showing the alignment of home location inference with statistical data. \par

Further, we assess the identified workplace locations to evaluate the generative model's performance. 
Figures \ref{fig:primary_activities}c and \ref{fig:primary_activities}d present the spatial distribution of workplace locations compared to the number of employees at 250m grids in central Stockholm and Gothenburg, respectively, the two most populated cities in Sweden. 
High employee density zones show a dense concentration of identified work locations, affirming the model’s ability to capture employment hubs.

\FloatBarrier

\subsection{Daily activity sequences}

\begin{figure}[htbp!]
    \centering
    \includegraphics[trim={0.0cm 0.0cm 0.0cm 0.8cm}, clip, width=.95\textwidth]{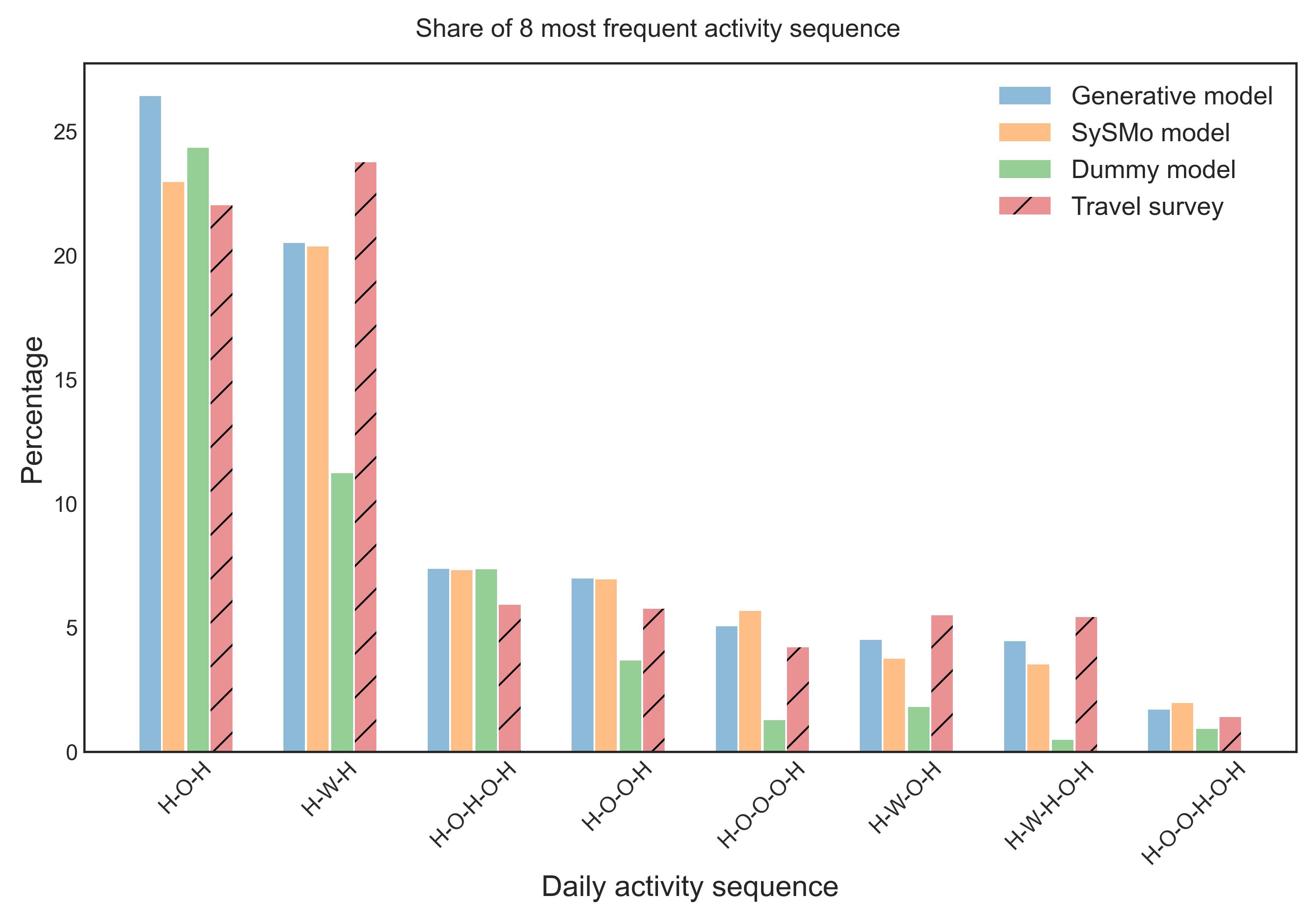}
    \caption{Share of eight most frequent daily activity sequences. \textmd{Home: H, Work: W, Other: O. These sequences comprise over 75\% of all sequences by the generative model results.}}
    \label{fig: act_sequence}
\end{figure}

Activity sequences show the order of activities during an average day, e.g., home-work-home (H-W-H). 
\Cref{fig: act_sequence} illustrates the distribution of the eight most common daily activity sequences, which account for over 75\% of all sequences generated by the generative model, and compares it with the other models and the travel survey. \par

The generative model produces a similar activity sequence distribution to the Synthetic Sweden Mobility (SySMo) model and the travel survey data. 
However, it slightly overestimates the Home-Other-Home sequence.
This discrepancy is likely due to the mobile phone data, which provides a more comprehensive view of secondary activities that can be underreported in survey data \citep{lee2016evaluation}. \par

In contrast, the dummy model does not fully capture the distribution of the sequences compared to the Synthetic Sweden Mobility (SySMo) model and the travel survey data. 
It has a low share of the Home-Work-Home sequence, one of the dominant activity sequences, and other less frequent sequences, e.g., Home-Work-Home-Other-Home. 
This discrepancy in infrequent sequences highlights the dummy model's limitations in modelling complex activity patterns. \par

To quantify the similarity in distribution across the models, we use Jensen–Shannon (JS) distances \citep{endres2003new}, ranging from 0 to 1, where 0 indicates identical distributions. 
The JS distance between the generative model and the SySMo model is 0.12. 
Similarly, the distance between the generative model and the travel survey is 0.07.
Both distances confirm close alignment in their sequence distributions. 
This contrasts with the dummy model's worse performance (JS distance= 0.41), indicating a lower level of alignment. 
This substantial difference suggests that the generative model's capability to synthesizing activity sequences is notably superior compared to the dummy model.
In other words, relying solely on the mobile phone geolocation data is insufficient to produce reasonable and useful activity sequences. \par

\begin{figure}[htbp!]
    \centering
    \includegraphics[trim={0.0cm 0.0cm 0.0cm 0.0cm}, clip, width=1.0\textwidth]{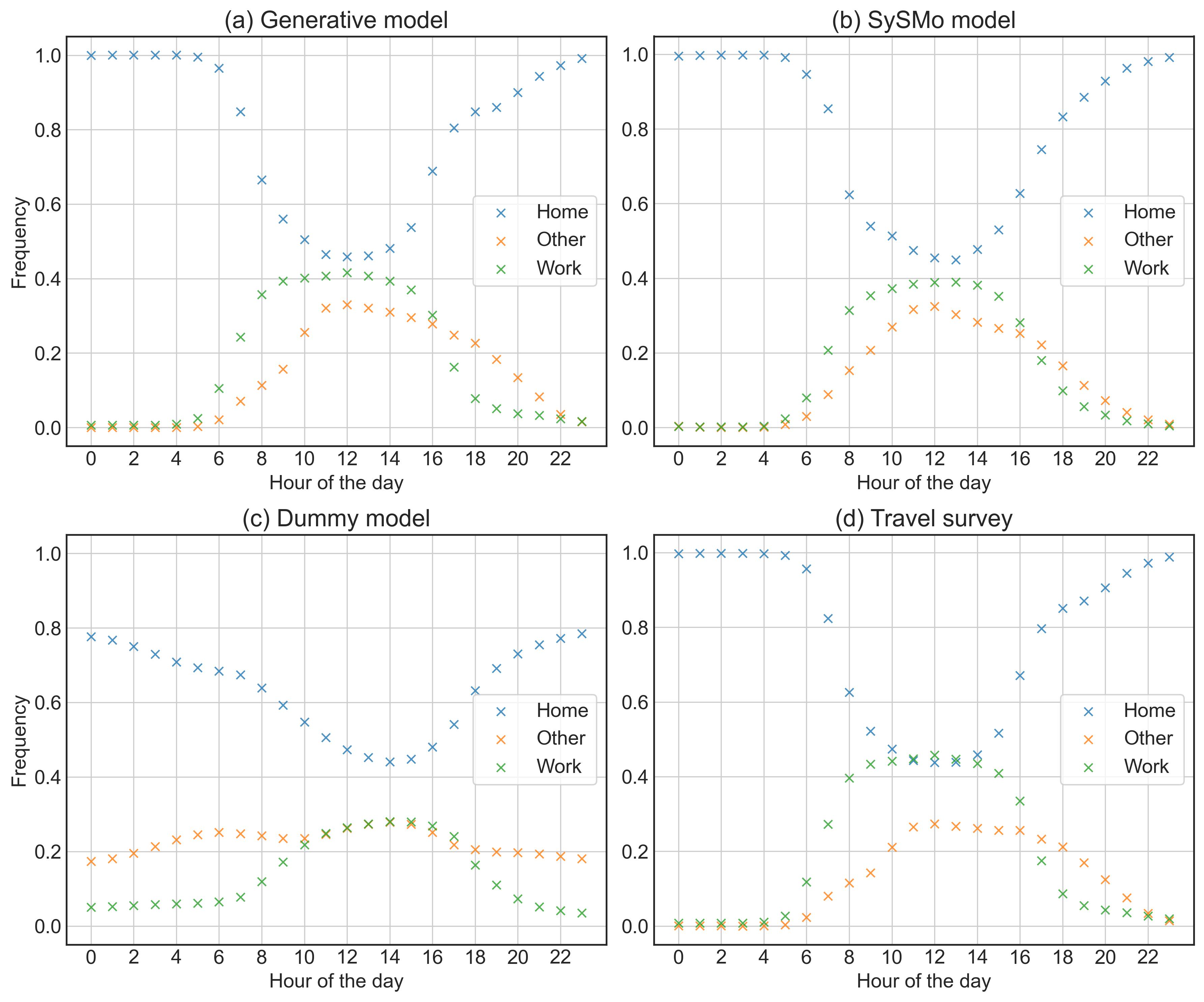}
    \caption{Aggregated temporal activity engagement patterns by activity type. Frequency is the share of individuals being at a certain activity.\textmd{(a) Generative model, (b) SySMo model, (c) Dummy model, (d) Travel survey.} }
    \label{fig: act_engagement}
\end{figure}

\begin{figure}[htbp!]
    \centering
    \newlength{\colwidth}
    \setlength{\colwidth}{0.333\textwidth}  
    
    \begin{minipage}{\colwidth}
        \centering
        \textbf{Home}
    \end{minipage}%
    \begin{minipage}{\colwidth}
        \centering
        \textbf{Work}
    \end{minipage}%
    \begin{minipage}{\colwidth}
        \centering
        \textbf{Other}
    \end{minipage}

    \newlength{\imgheight}
    \settoheight{\imgheight}{\includegraphics[width=1.0\textwidth]{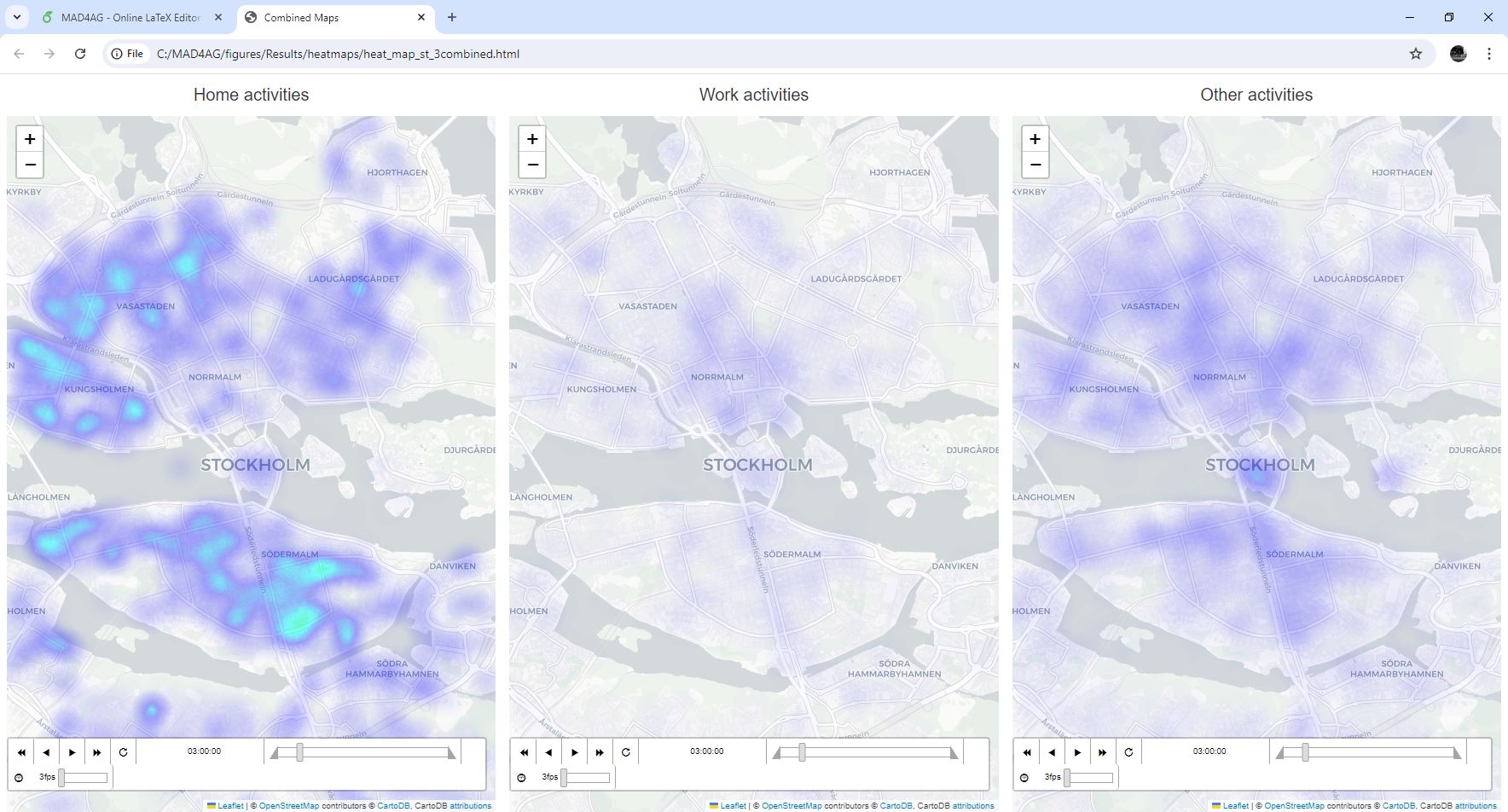}}
    \begin{tikzpicture}
        \node[rotate=90, anchor=south] at (-7.43, 0) {\textbf{03:00 am}};
        \node[inner sep=0] at (0,0) {\includegraphics[trim={0.0cm 4cm 0.0cm 5.5cm}, clip, width=1.0\textwidth]{heat_map_st_3combined_3.JPG}};
    \end{tikzpicture}\\
    \begin{tikzpicture}
        \node[rotate=90, anchor=south] at (-7.43, 0) {\textbf{09:00 am}};
        \node[inner sep=0] at (0,0) {\includegraphics[trim={0.0cm 4cm 0.0cm 5.5cm}, clip, width=1.0\textwidth]{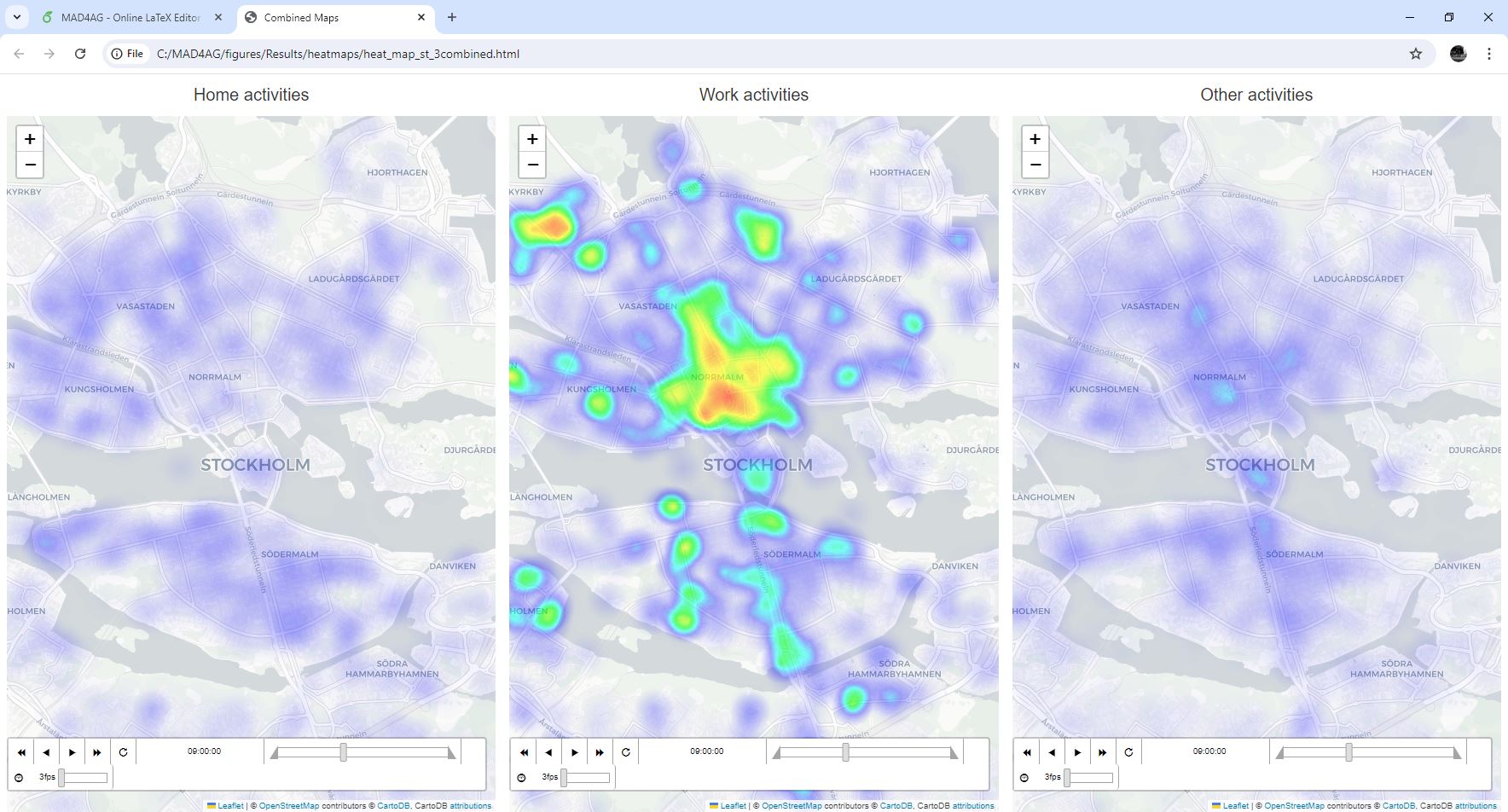}};
    \end{tikzpicture}\\
    \begin{tikzpicture}
        \node[rotate=90, anchor=south] at (-7.36, 0) {\textbf{03:00 pm}};
        \node[inner sep=0] at (0,0) {\includegraphics[trim={0.0cm 4cm 0.0cm 5.5cm}, clip, width=1.0\textwidth]{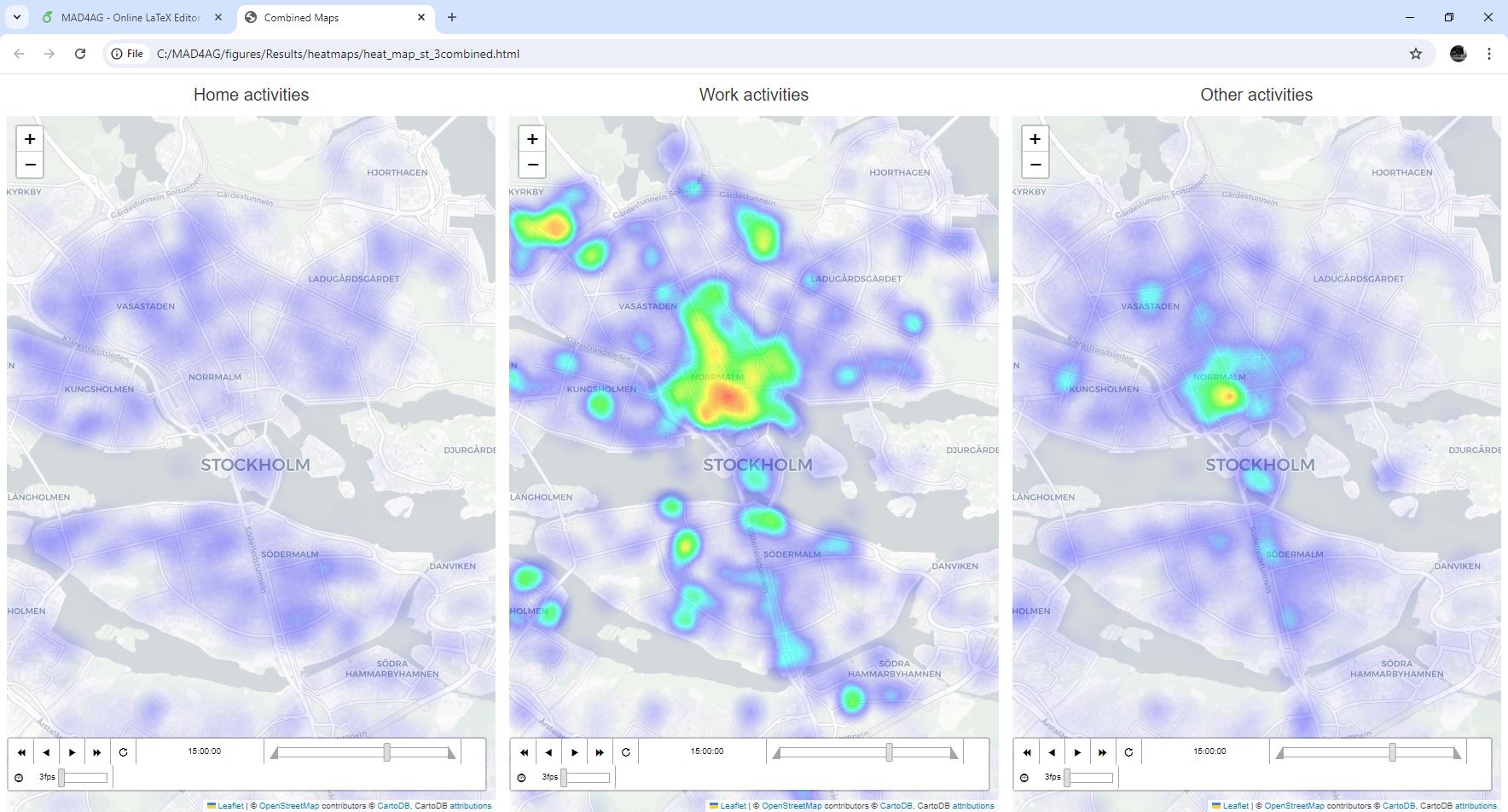}};
    \end{tikzpicture}\\
    \begin{tikzpicture}
        \node[rotate=90, anchor=south] at (-7.36, 0) {\textbf{09:00 pm}};
        \node[inner sep=0] at (0,0) {\includegraphics[trim={0.0cm 4cm 0.0cm 5.5cm}, clip, width=1.0\textwidth]{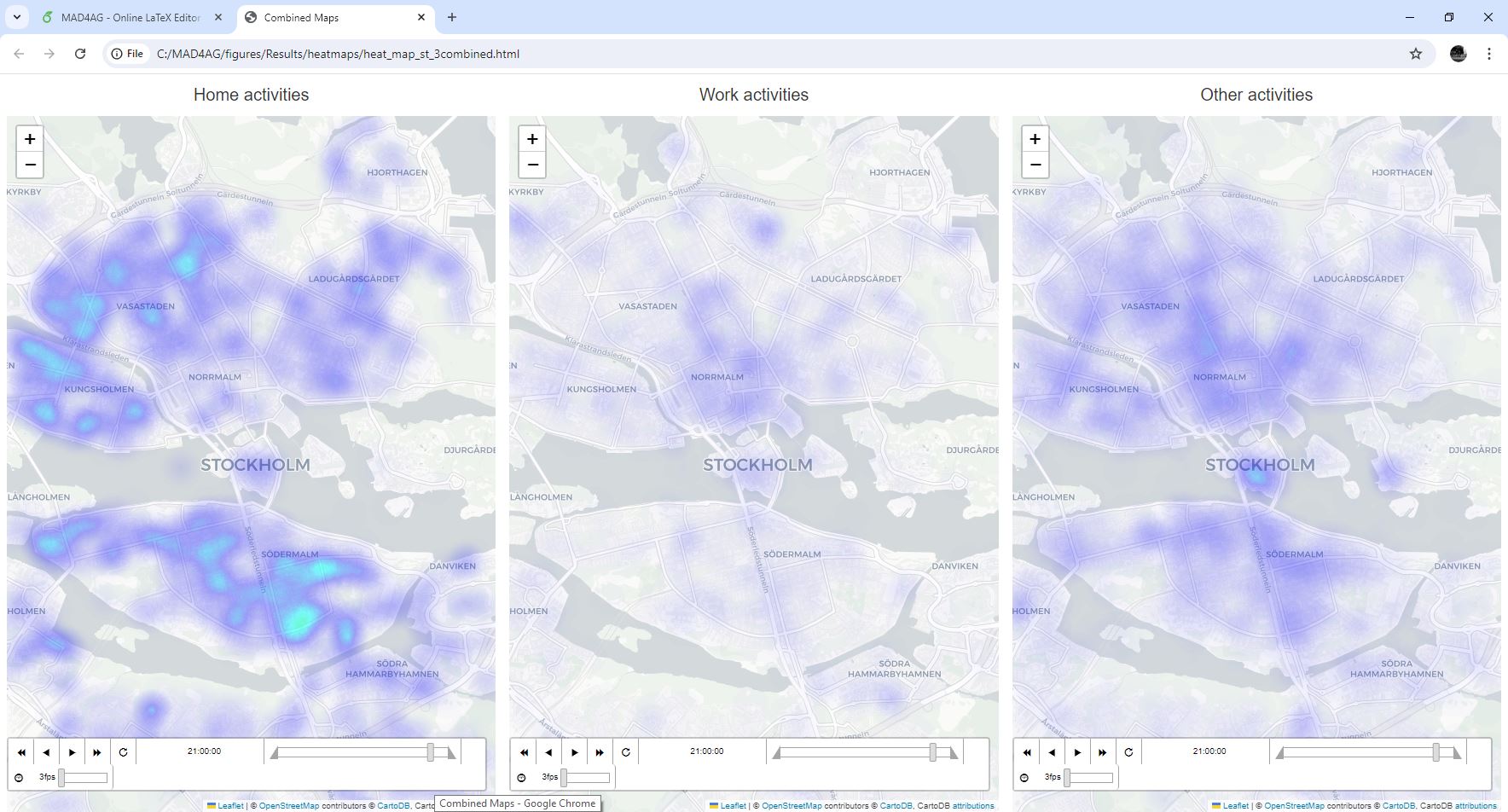}};
    \end{tikzpicture}\\
    \includegraphics[trim={0.0cm 0.3cm 0.25cm 0.0cm}, clip, width=0.92\textwidth]{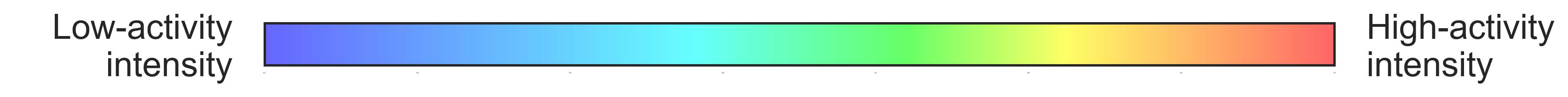}\\
    \caption{Spatio-temporal distribution of activity participation by type in central Stockholm. \textmd{The warm colours indicate hotspots of high activity intensity, while cold colours indicate relatively low activity intensity for a certain activity type.}}
    \label{fig: heatmap}
\end{figure}

\subsection{Temporal activity patterns}

We assess hourly activity patterns to better understand temporal changes, which are vital for effective scheduling.
Figure \ref{fig: act_engagement} shows the temporal distribution of activity participation by activity types from midnight to the following day, comparing results across different models and the travel survey. 
The generative model follows the home activity patterns compared to the SySMo model and the travel survey data during both night and daytime hours. 
It also has work and other activity engagement patterns similar to SySMo and the survey. \par

The dummy model captures the general pattern of hourly activity participation but underrepresents home activity participation during the night hours.
Moreover, compared to the SySMo model and travel survey data, it overrepresents other daytime activities. 
These discrepancies underscore the dummy model’s limitations and, conversely, demonstrate the generative model’s ability to capture both nighttime and daytime activity patterns accurately. \par

The spatio-temporal activity engagement patterns by type in central Stockholm (Figure \ref{fig: heatmap}) corroborate with the above observations. 
Home activities are prevalent throughout the day in residential areas, increasing during night hours.
Work activities peak during daytime hours, particularly in central business districts, e.g., Norrmalm and Östermalm. 
While the distribution of other activities is geographically similar to work activities, their occurrence extends into night hours, illustrating a distinct temporal variation. Figure \ref{fig:geolocationsbyactivities} in the appendix shows the activities' geolocations by the time of day in central Stockholm.

\FloatBarrier

\subsection{Trip distances}

We evaluate the generative model's travel patterns by comparing spherical trip distances between activity locations across different models and travel survey data. 
This analysis presents the model's ability to simulate daily trips. \par

\begin{table}[htbp!]
\centering
\caption{Trip distances comparison by models (in km)}
\begin{tabular}{llccc}
\hline
\textbf{Model}   & \textbf{Trip type}     & \textbf{Mean}    & \textbf{Median}  & \textbf{90th percentile} \\
\hline
Generative model& Overall & 21.5 & 3.7 & 56.3 \\
SySMo model& Overall & 9.8 & 3.4 & 21.3  \\
Dummy model& Overall & 35.3 & 6.5 & 124.6 \\ 
Travel survey& Overall & 12.5 & 4.3 & 24.1 \\ 
Generative model& Commuting & 18.3 & 5.3 & 39.7 \\
SySMo model& Commuting & 11.7 & 5.6 & 27.6 \\
Dummy model& Commuting & 32.9 & 8.4 & 104.4 \\
Travel survey& Commuting & 15.5 & 7.9 & 30.24 \\
\hline
\end{tabular}

\label{tab:model_comparison}
\end{table}

\begin{figure}[htbp!]
    \centering
    \includegraphics[trim={0.0cm 0.2cm 0.0cm 0.9cm}, clip, width=1.0\textwidth]{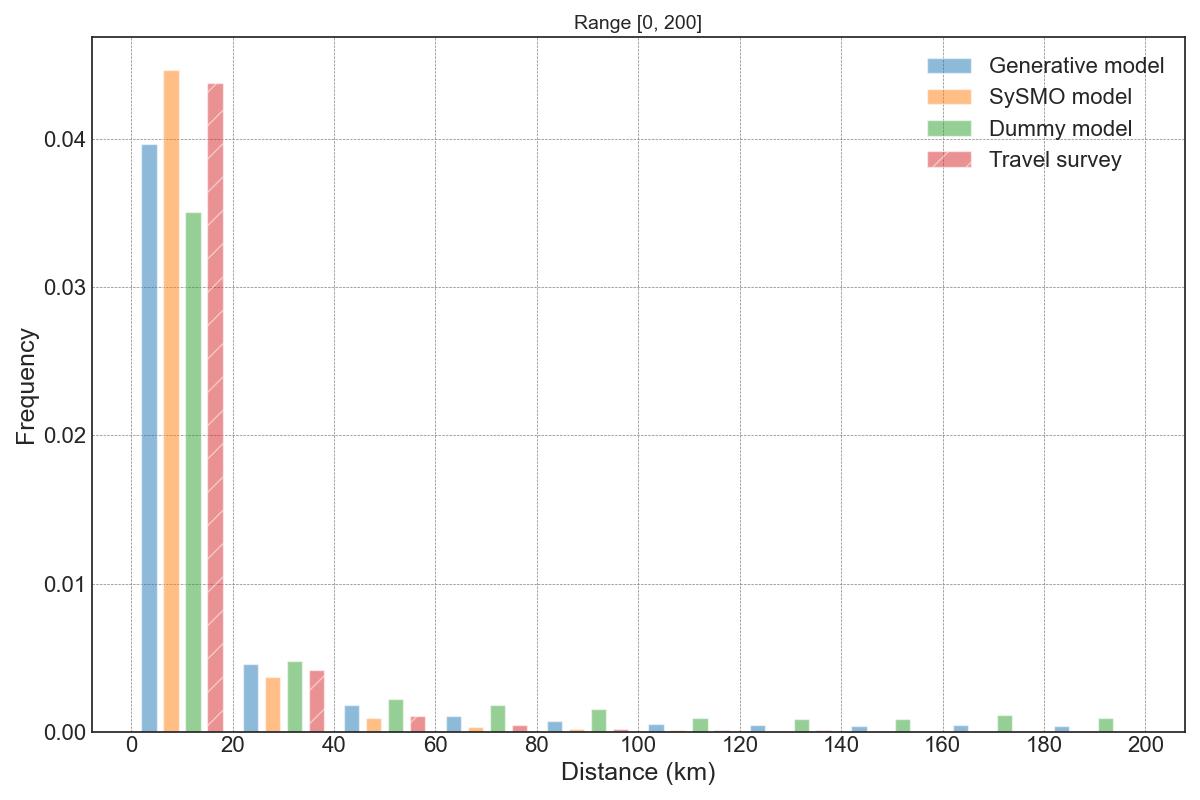}
    \caption{Trip distance distribution by models \textmd{(in the range of 0 to 200 km)}.}
    \label{fig: trip_dist}
\end{figure}

The generative model captures the trip distribution patterns observed in the SySMo model and the travel survey.
Figure \ref{fig: trip_dist} illustrates the trip distance distribution, and Table \ref{tab:model_comparison} compares the mean, median, and 90th percentile values of trip distances across the models.
The generative model shows slightly lower frequency values in the 0-20 km range but tends to overestimate distances for trips exceeding 60 km.
It also reports the mean and 90th percentile trip distances substantially higher than the travel survey and SySMo model, reflecting its inclusion of longer trips. 
Traditional travel surveys often under-report long-distance trips \citep{janzen2018closer}.
However, with a median trip distance of 3.9 km, the generative model closely aligns with the data being compared, indicating its capture of central trip distances despite the mobile data including a broader range of trips.
The mean and median commuting trip distances of the generative model also align well with the SySMo model and the travel survey.
In contrast, the dummy model consistently shows the highest values, largely diverging from the survey data and SySMo model. \par

\FloatBarrier

\subsection{Robustness and variance}

The generative model creates multiple average weekday simulations by iteratively finding twin travellers and activity locations for each individual in the mobile phone dataset. 
This produces variance in the resulting activity-travel plans while maintaining activity patterns at the aggregate level. \par

\begin{figure}[htbp!]
    \centering
    \begin{tabular}{lr}

        \begin{minipage}{0.5\textwidth}
            \centering
            \includegraphics[trim={0.0cm 0.0cm 0.0cm 0.0cm}, clip, width=1\textwidth]{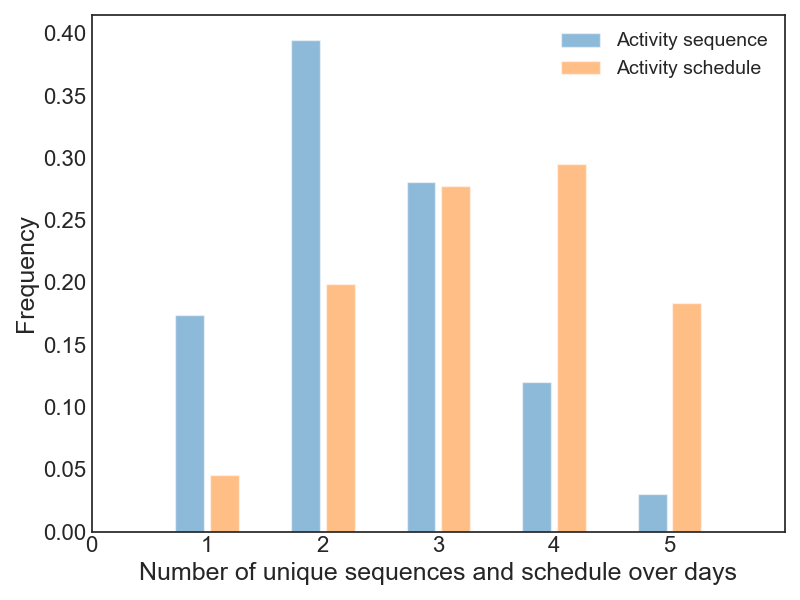} \\
            (a) 
        \end{minipage}
        &
        \begin{minipage}{0.5\textwidth}
            \centering
            \includegraphics[trim={0.3cm 0.2cm 0.9cm 1.5cm}, clip, width=1.03\textwidth]{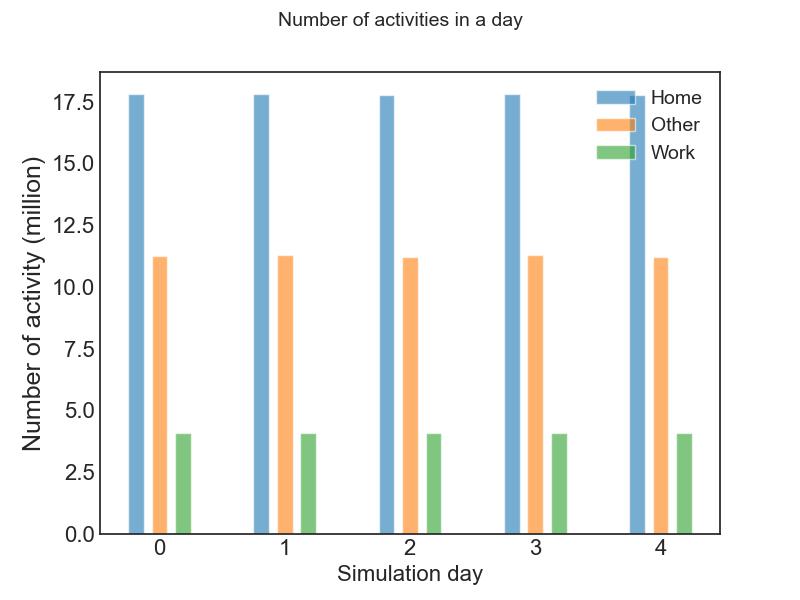} \\
            (b) 
        \end{minipage}
    \end{tabular}
    \caption{Comparison of multiple average weekday simulations: \textmd{(a) the number of unique activity sequences and schedules over five simulation days, (b) Aggregated activity engagement by activity types across five simulation days.}}
    \label{fig: multiday_sim}
\end{figure}

First, the model introduces variations in all components of the activity schedules, including activities' sequence, types, start/end times, and other activity locations while preserving the residential and workplace locations. 
Figure \ref{fig: multiday_sim}a illustrates the diversity in activity sequences and schedules over five simulation days. 
While more than half of the individuals have the same or only two different activity sequences throughout the simulation period, their activity sequences show variation because of the activities' start/end time and the location of other activities. 
This variation adds depth to the simulation by reflecting realistic daily dynamics, leveraging abundant geolocations for the other activities from mobile phone data. \par

Second, the model produces stable aggregated activity patterns despite the individual variance.
Figure \ref{fig: multiday_sim}b compares aggregated activity engagement by activity types across five simulation days. 
Despite the observed variations at the individual level, there is minimal fluctuation in overall activity participation across the simulations, indicating consistent behaviour patterns among the simulated population. \par

\section{Discussion and Concluding Remarks}\label{sec: discussion}

This study introduces a generative model that synthesises individuals' daily activity-travel patterns from extensive mobile phone GPS records. Mobile phone application data, a relatively new big data source, offers the potential for generating detailed activity-travel patterns, but its individual-level sparsity complicates creating realistic activity plans. Previous studies have highlighted the benefits of similar big data sources in activity-based modelling studies but have not adequately addressed data sparsity and population bias issues \citep{alexander2015origin, jiang2017activity, grujic2022combining}. By combining mobile data with a travel survey, our study addresses mobile data's sparsity and biases and realistically simulates average weekday activity schedules for 263,000 individuals in Sweden.

To assess the generative model's performance, we compare its results with those from the underlying travel survey data, Synthetic Sweden Mobility (SySMo) model, and the dummy model that relies solely on mobile phone data. The generative model's activity pattern closely aligns with the underlying survey data and the SySMo model, substantially outperforming the dummy model. The generative model's activity sequence distribution achieves a Jensen–Shannon distance of 0.07 compared to the survey and 0.12 compared to the SySMo model, indicating high similarity. Additionally, the comparison of hourly activity participation during the day further confirms the close alignment of the generative model with the survey data and the SySMo model. Meanwhile, the dummy model deviates from the realistic pattern revealed by the other models. The generative model accurately captures trip distance distributions in travel patterns, with median trip distances aligning with the survey data and SySMo model. The dummy model shows the highest values, diverging from both the survey and SySMo models.

Furthermore, our study advances the state-of-the-art primary activity identification approaches by introducing a temporal-score approach. In anonymised mobile phone data, these locations are typically inferred using temporal rules (e.g., the studies by \citet{chen2014traces, sadeghinasr2019estimating}), often failing to account for individual differences in activity participation patterns. Our approach considers variations in activity engagement at different observation times. The inferred home and work locations align well with the grid-level population and employee statistics, demonstrating the model's capability to accurately identify primary activities.

While the generative model effectively generates activity-travel patterns from mobile data, it is also important to acknowledge the model's limitations and consider the broader context of activity-based models. One major challenge lies in the disaggregated evaluations of the activity travel schedules. Most models in the literature lack validation due to the absence of ground truth data. Despite this study's promising findings, this gap limits our ability to test the model's reliability and accuracy in a precise way.
Furthermore, the combined data sources have different collection periods: survey data from 2011 to 2016 and mobile application data from June to December 2019. Future work should carefully align the data sources and explore any possible travel behaviour changes between years in Sweden.

Another limitation is that the model captures variability in travel behaviour by generating multiple plans but only for an average weekday. This means that the model does not account for weekly variations of activity demand. Analysing activity patterns over extended periods, such as a week, could provide valuable insights into infrastructure measures and travel behaviour stability and variability \citep{moeckel2024activity}. Future research aims to address this limitation by extending the model to weekly activity pattern simulation.

In this study, we utilise GDPR-compliant data obtained from mobile applications with explicit consent from device users to share their locations. Data privacy is a critical concern, and even anonymised data can pose risks of revealing individual identities. To mitigate privacy risks, the study strictly avoids publishing trajectories that could be traced back to specific individuals. The proposed model also holds the potential to provide representative, synthetic, and privacy-preserving mobility data.

The proposed model's methodology demonstrates strong transferability by leveraging large-scale geodata and survey data. While the model gives promising results in Sweden, applying the model in diverse settings requires fine-tuning parameters to account for variations in data collection, infrastructure, and behaviour. Adaptations to local conditions and thorough validation are essential to ensure relevance and utility.


\vspace{10cm}

\section*{Ethics statements}
The authors declare that this work does not involve the use of social media data or experimentation with animals. \par

We followed the Chalmers Institutional Review Board (IRB) guidelines in line with the Swedish Ethical Review Act (2003:460) for research involving humans and the General Data Protection Regulation (GDPR) 2016/679. Given the nature of the data analyzed, the study was exempt from ethical review under the Swedish Ethical Review Act.

\section*{CRediT Author Statement}

Conceptualization: Çağlar Tozluoğlu (Ç.T.), Yuan Liao (Y.L.), Frances Sprei (F.S.); 
methodology: Ç.T., Y.L., F.S.;
software: Ç.T., Y.L.;
validation: Ç.T.;
data curation: Ç.T., Y.L.;
writing - original draft: Ç.T.;
writing - review \& editing: Ç.T., Y.L., F.S.;
project administration: F.S.

\subsection*{Acknowledgments} 
This research is funded by the Swedish Research Council Formas (Project Number 2018-01768). 
The authors acknowledge Sonia Yeh for her intellectual contributions to the study's methodology. 
Additionally, the authors sincerely thank Jorge Gil for providing the mobile phone application data and Mattias Rydström and Diana Salim for their master thesis as valuable prior work.

\subsection*{Declaration of interests} 

The authors declare that they have no known competing financial interests or personal relationships that could have appeared to influence the work reported in this paper. 

\subsection*{Declaration of generative AI and AI-assisted technologies in the writing process}

During the preparation of this work, the authors used ChatGPT in order to improve language and readability, with caution. After using this tool/service, the authors reviewed and edited the content as needed and take full responsibility for the content of the publication.

\vspace{6cm}

\bibliography{sample}

\appendix
\section{Appendix}
\renewcommand{\thefigure}{A.\arabic{figure}}
\renewcommand{\thetable}{A.\arabic{table}}
\setcounter{figure}{0}
\setcounter{table}{0}

\subsection{Data}\label{seca:data}

The proposed model combines two data sets for synthesising activity-travel plans: the mobile application data and the Swedish national travel survey (2011--2016) \citep{RVU}. We also use Demographic statistics areas (DeSO zones) \citep{DeSO}, and the building locations \citep{buildings} as auxiliary data. Below is a brief description of each data set.

\subsubsection{Mobile application data}
\label{data:mobile}
Mobile application data consists of anonymised geolocation information collected from location-enabled applications on individuals' smartphones, formatted as (id, latitude, longitude, time).
This kind of data features extensive spatial and population coverage at a relatively low cost. 
Although they provide mobility traces from a large amount of devices, they generally lack socio-demographic information about the device users due to privacy concerns. \par

\begin{figure}[h!]
    \centering
    \includegraphics[trim={0.0cm 0.0cm 0.0cm 0.0cm}, clip, width=1.0\textwidth]{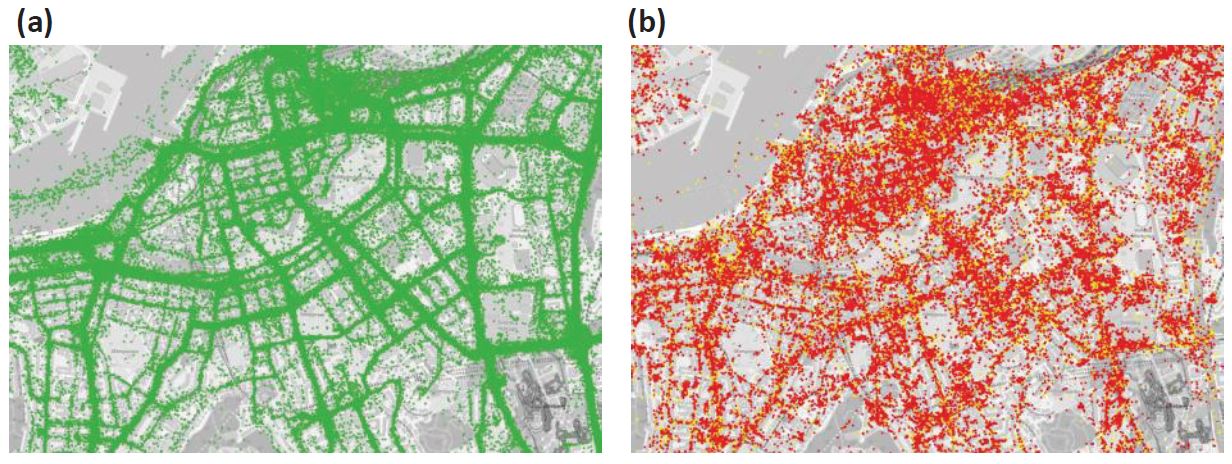}
    \caption{Geolocations from mobile application data in Gothenburg \citep{liao2024uneven}. \textmd{(a) Movement points. (b) Stationary points.}}
    \label{fig: geolocations}
\end{figure}

This study uses a dataset sourced from a diverse set of mobile apps used by adult smartphone users in Sweden\footnote{http://www.pickwell.co/}.
The data covers seven months in 2019 (from June to December) with approximately 25 million daily GPS records.
In the raw data, we have around 4.6 billion records from 1 million devices. 
Mobile application data captures geolocations of individuals in motion (\Cref{fig: geolocations}a) and stationary (\Cref{fig: geolocations}b) at the moment when they interact with certain phone applications. \par

However, the mobile application geolocations have varying sampling frequency across users, periods, and geographic regions. 
This variability is primarily due to users' preference for using certain geolocation-enabled applications across different places at different times. 
This intermittent data collection leads to sparsity in observation and also introduces sampling bias among users, complicating the analysis of human mobility patterns.

\subsubsection{Swedish National Travel survey}
\label{data:survey}

The Swedish National Travel Survey \citep{RVU} collects self-reported, one-day travel diaries from individuals with their socio-demographic backgrounds. 
The survey period is between 2011 and 2016, with around 38,000 participants aged 6--84 years. For our study, we only use weekday activity-travel patterns, excluding data from holidays and weekends. \par

The data preprocessing involves removing records with incomplete or inaccurate trip details, e.g., trip origin and destination information. 
This step ensures that only participants with complete activity-travel patterns for a weekday are included in our analysis. 
To have a similar population with the mobile application data, we also exclude respondents below 18. 
After preprocessing, the data contains approximately 15,000 individuals for weekdays. 
Although the preprocessing step results in a reduced sample size and a further sparsity in the survey's geographic coverage across Sweden, the data offers a detailed snapshot of individuals' daily activities, providing valuable insights into their routines and travel behaviour. \par

\subsubsection{Demographic statistics areas (DeSO zones)}
\label{data:deso}

Demographic statistical areas (DeSO zones) \citep{DeSO} obtained from Statistics Sweden are a nationwide division. 
DeSO zones are formed by dividing municipal boundaries into smaller areas, each with a population minimum of 700 and a maximum of 2,700 inhabitants. 
Across Sweden, there are a total of 5,984 DeSO zones. 
For each zone, the data contains various statistics, i.e., the population size, the population size by age groups, the number of employees and students, etc. 
We use statistics to enhance the representativeness of the results.

\subsubsection{Building data}

Building data \citep{buildings} derived from property registers provide information on more than 8.6 million buildings across Sweden. 
This comprehensive dataset includes building locations, footprints, and usage types (detached houses, apartments, workplaces, schools, etc.). 
Due to factors such as GPS inaccuracies and user movement, geolocations obtained from the Mobile Application Data (MAD) can be dispersed around actual activity locations. 
We employ building data to determine the precise activity locations.

\subsection{Stop detection}\label{seca:stop_detection}

We detect stops from individuals' geolocation trajectories, represented as ($id, lat, lon, time$). 
Stop refers to a location where an individual stays for a certain duration. 
These stops are critical for analysing individuals' activity-travel patterns. 
Various algorithms are available for stop detection, e.g., infostop \citep{aslak2020infostop}, DBSCAN \citep{ester1996density}, and scikit-mobility \citep{JSSv103i04}. 
For this study, we follow a previous study \citep{liao2024uneven} using the infostop algorithm due to its resilience to noise in data, efficiency with extensive data, and capability to conduct analyses across multiple users simultaneously \citep{aslak2020infostop}.

The infostop algorithm operates with parameters, $r_1, r_2, t_{min}, t_{max}$. The algorithm begins by detecting stops when an individual remains within a distance $r_1$ for a minimum duration $t_{min}$ and a maximum time difference $t_{max}$ between consecutive records. Subsequently, the algorithm clusters these stops spatially, with a parameter $r_2$. Each detected stop is recorded as a tuple ($id, lat, lon, start, end $). Table \ref{tab:infostop} shows the employed parameters in the infostop algorithm.

\begin{table}[htbp!]
\small
    \centering
    \caption{The infostop algorithm parameters and set values \citep{liao2024uneven}.}
    \begin{tabular}{llll}
    \hline
    Parameter & Definition & Unit & Value \\ \hline
    $r_1$ & \begin{tabular}[c]{@{}l@{}}The maximum allowable distance for roaming between two points\\in the same stop  \end{tabular} & meter & 30 \\
    $r_2$ & The maximum distance between two stops & meter & 30  \\
    $t_{min}$ & The minimum duration of a stay & min & 15  \\
    $t_{max}$ &  \begin{tabular}[c]{@{}l@{}}The maximum time difference between two consecutive records\\ in the same stop \end{tabular} & hour & 3  \\
    \hline
    \end{tabular}
    \label{tab:infostop}
\end{table}

\subsection{Temporal-score approach for inferring primary activities}\label{seca:home-work_detection}
We design a scoring approach to infer primary activity locations, integrating the temporal rules with the empirical evidence into the methodology.
This approach considers variations in activity patterns at different observation times rather than treating all hours equally while assessing the likelihood of locations.

\begin{figure}[htbp!]
    \centering
    \includegraphics[trim={0.0cm 0.0cm 0.0cm 0.6cm}, clip, width=0.95\textwidth]{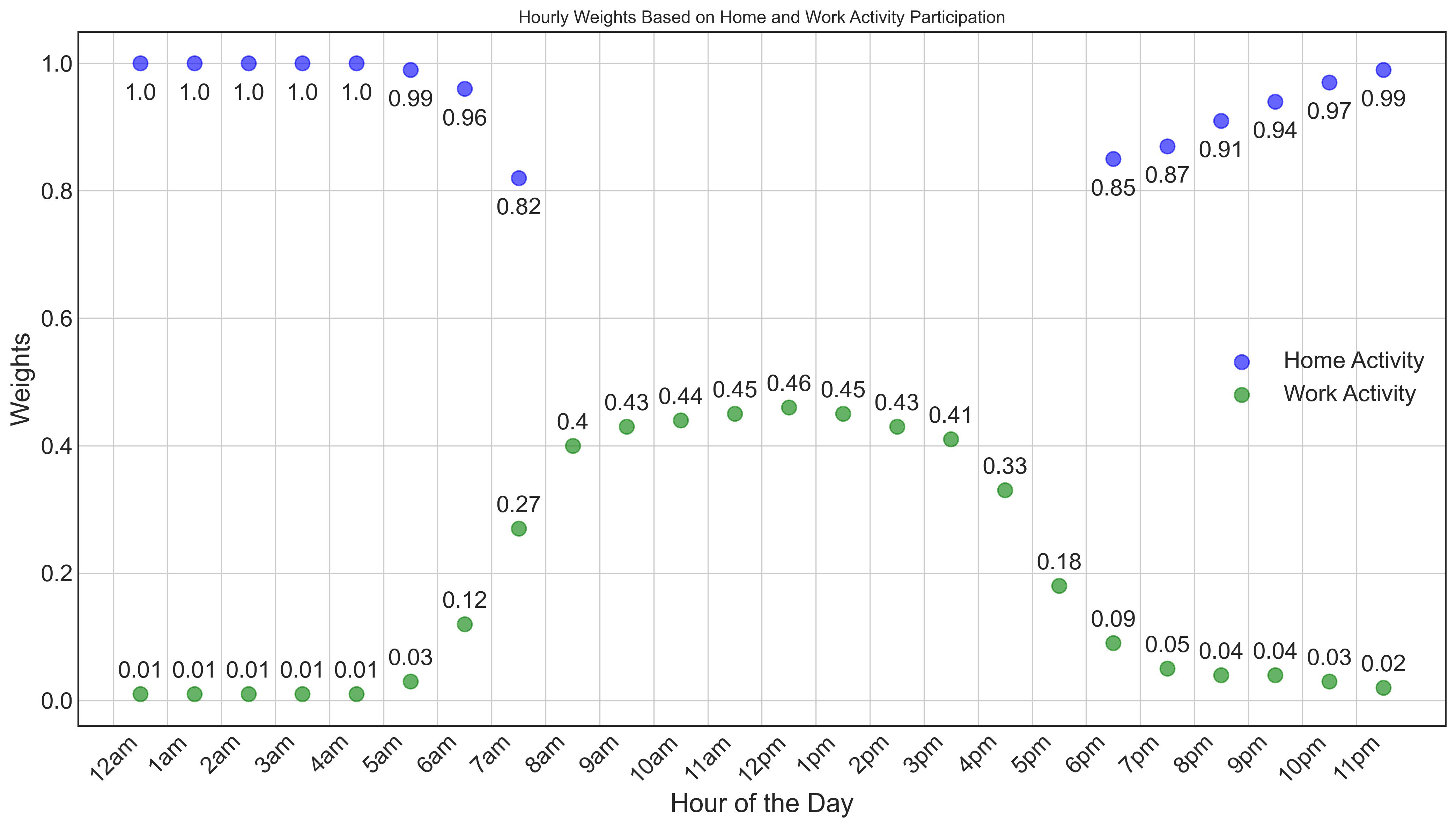}
    \caption{Hourly weights based on activity participation in the travel survey. \textmd{The sum of weights can exceed one due to rounding errors or some individuals engaging in multiple activities within an hour.}}
    \label{fig: weights}
\end{figure}

We first calculate weights using travel survey data to deduce temporal rules.
These weights represent the ratio of hourly activity participation for a specific activity type (e.g., home or work) to the overall activity participation in the travel survey. 
For home activities, weights are assigned during predefined nighttime hours from 6:00 pm to 7:59 am. 
For work activities, we extend weight calculations across all hours of the day to accommodate varying work schedules, including late hours, thereby offering more flexibility. 
Figure \ref{fig: weights} illustrates these weights.

We then calculate a score for each activity location using the activity records with the stay duration information. 
Individuals can have abundant activity records in activity locations. 
To determine the score, each hour spent at a location is multiplied by its corresponding hourly weight. 
We then sum up these weighted hours and calculate the location's score. 
For example, if an individual has two activity records at a location from 9:00 pm to 5:59 am and from 7:00 pm to 3:59 am, each hour spent during these periods is weighted according to the specific hourly weights and summed to produce the location's overall score.

Let $W=(w_1,w_2,w_3,...,w_{24} )$ represent the hourly weights where each $w_i$ is associated with a specific hour of the day. 
For a given activity location $A$, let $A=(a_1,a_2,a_3,..., a_i)$ denote the set of activity records at this location, with each record $a_i$ containing the hours $h_i$ spent during each period of activity. 
The score $S_A$ for location $A$ can be calculated as follows:

\begin{align}
    S_A = \sum_{i} a_i \sum_{j} w_j \cdot h_j 
    \label{eq:score}
\end{align}

\subsection{Searching twin travellers}\label{seca:search_twins}

To find twin travellers for the individuals in the mobile phone dataset, we first divide individuals in the mobile phone data and the travel survey data into groups who share similar demographic and activity-travel attributes. 
The variables used for grouping are detailed in Table \ref{tab: variables} and include residential region, urban density of residence, employment status, average trip distance, and commuting distance.

We define residential regions by the three greater regions in Sweden: Norrland, Svealand, and Götaland. 
Urban density, derived from DeSO zones, categorises residential locations as high or low. 
High urban density indicates central urban locations, while low urban density represents areas outside major population concentrations or urban areas \citep{DeSO}. 
Employment status has two categories: employed (Yes) and unemployed (No). 
For the travel survey data, we based on survey responses, while for the mobile phone data, employment status is inferred based on the presence of a workplace location.

We also categorise trip distances into two levels: low and high. 
An individual's average trip distance that exceeds the median of all average trip distances is classified as long; otherwise, it is classified as short. 
The same median-based classification applies to commuting distances.

In the mobile phone dataset, all attributes are applied directly to group individuals. 
However, given the limited sample size of the travel survey data, we implement a step-wise approach for grouping the travel survey participants to prevent overly sparse groups. 
We begin grouping individuals by region and then sequentially by urban density, employment status, average trip distance, and commuting distance. 
At each step, we evaluate the size of the groups, i.e., how many participants belong to each group. 
If a group includes at least 50 individuals, we further subdivide the group based on the next attribute in the sequence. 
This approach is aimed at having sufficient travellers in each group.\par

\begin{table}[htbp!]
\centering
\caption{Variables used for categorizing travellers (Section \ref{sec:cate_trav}). \textmd{Order column shows the variables' sequence in the step-wise grouping algorithm for the travel survey participants. $*$ Defined by the DeSo zones' statistics.}}
\label{tab: variables}
\begin{tabular}{lll}\hline
\textbf{Order} & \textbf{Variable} & \textbf{Levels}  \\  \hline
1 & Residential region &  Svealand \\
 & & Götaland \\
 & & Norrland \\ 
2 & Urban density of their residence$^*$ &  Low \\
 & & High \\ 
3 & Employment status &  Yes \\
 & & No \\ 
4 & Average trip distance & Short (< 4.3 km) \\
 & & Long ($\geq$ 4.3 km)  \\ 
5 & Commuting distance & Short  (< 7.9 km)\\
 & & Long ($\geq$ 7.9 km) \\ \hline
\end{tabular}
\end{table}

\subsection{Distribution of activities by time of day}\label{seca:geolocationsbyactivities}

\begin{figure}[htbp!]
    \centering
    \begin{tabular}{lr}
        \begin{minipage}{0.5\textwidth}
            \centering
            \includegraphics[trim={11cm 3cm 13cm 6cm}, clip, width=1.03\textwidth]{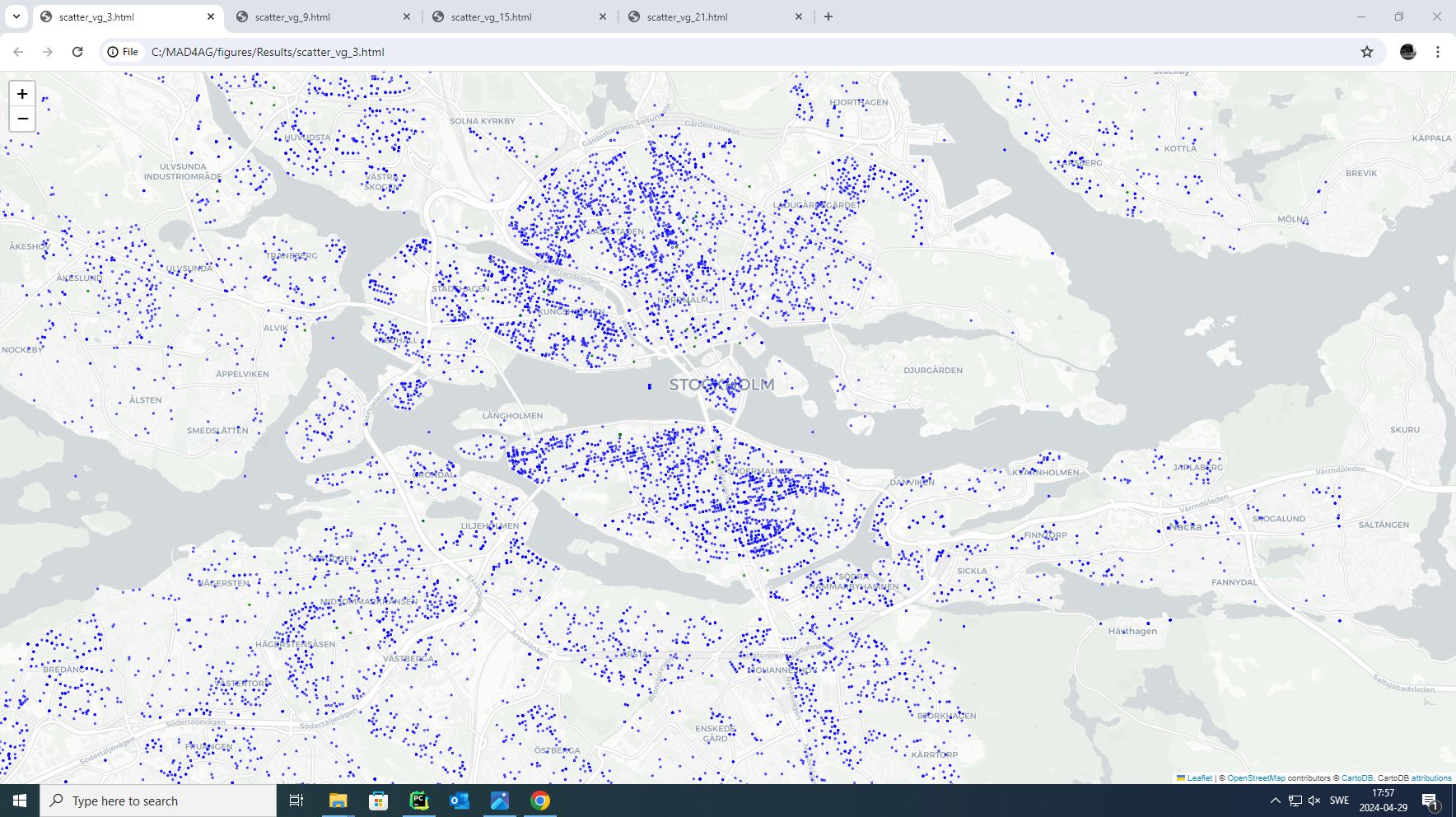} \\
            03:00 am \\[6pt]
            \begin{tikzpicture}
                \node[anchor=south west,inner sep=0] (image) at (0,0) {\includegraphics[trim={11cm 3cm 13cm 6cm}, clip, width=1.03\textwidth]{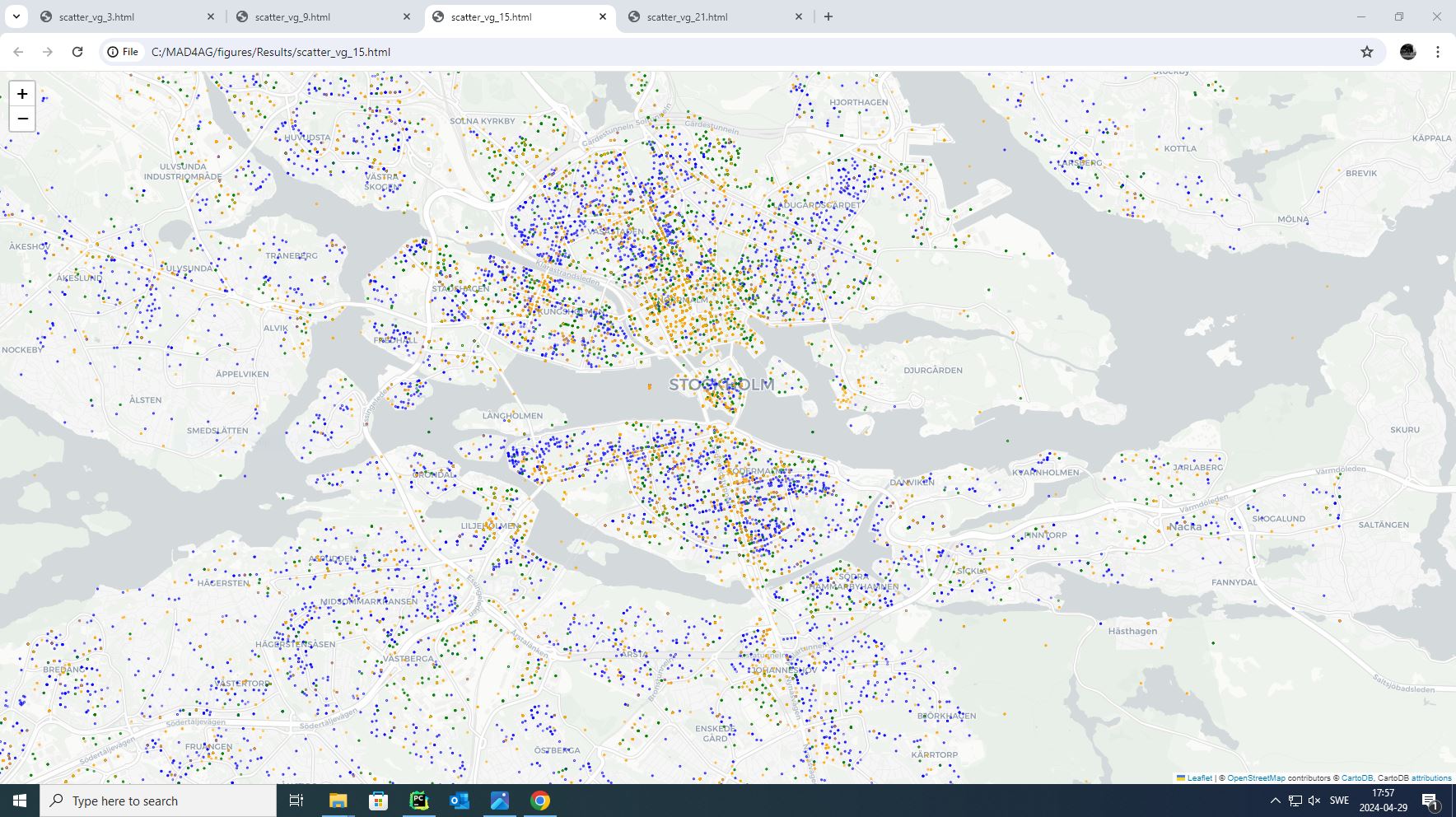}};
                \begin{scope}[x={(image.south east)},y={(image.south west)}]
                    \node[anchor=south west] at (0,1) {\includegraphics[width=0.3\textwidth]{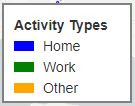}};  
                \end{scope}
            \end{tikzpicture}
            03:00 pm
        \end{minipage}
        &
        \begin{minipage}{0.5\textwidth}
            \centering
            \includegraphics[trim={11cm 3cm 13cm 6cm}, clip, width=1.03\textwidth]{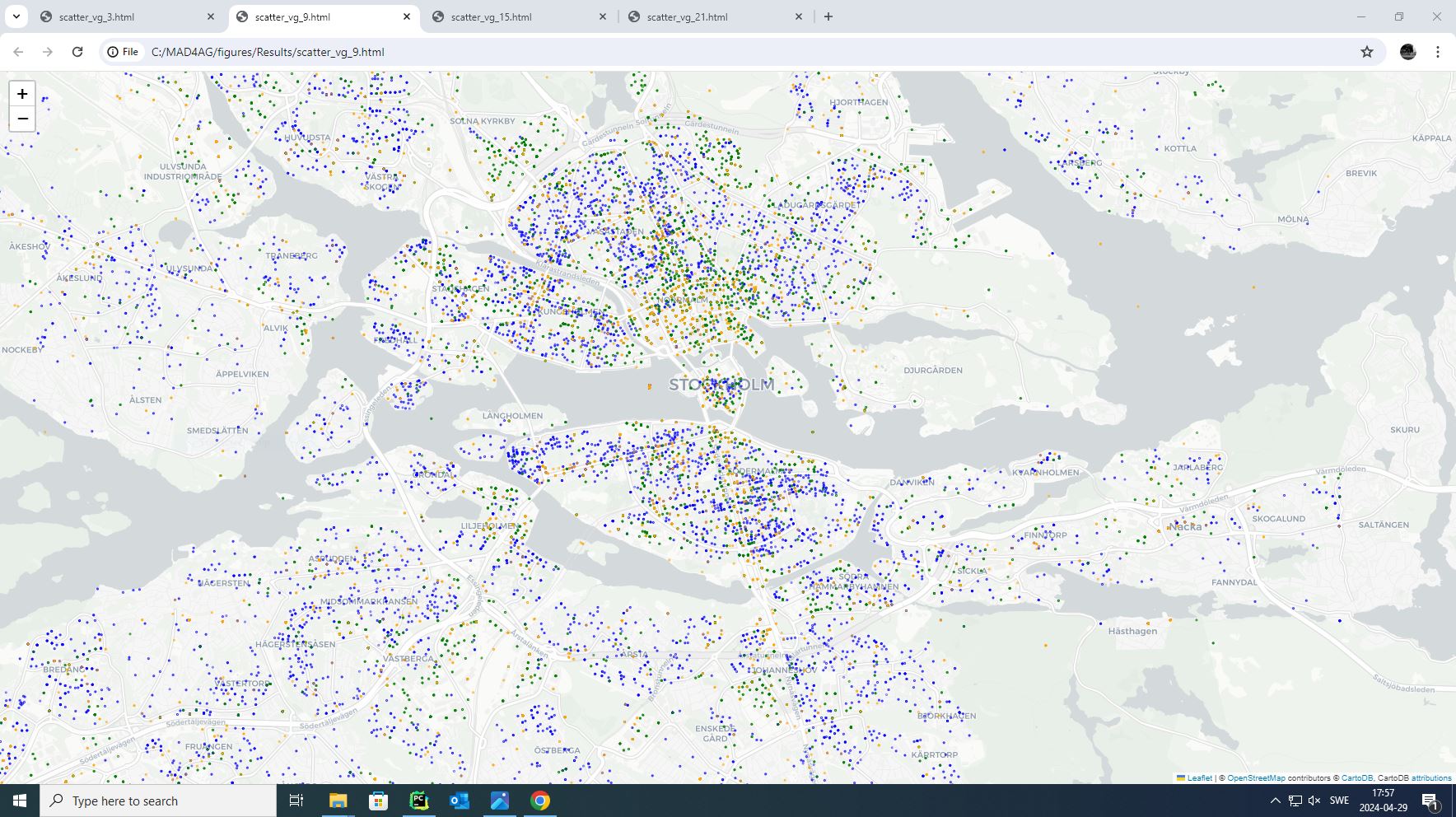} \\
            09:00 am \\[6pt]
            \includegraphics[trim={11cm 3cm 13cm 6cm}, clip, width=1.03\textwidth]{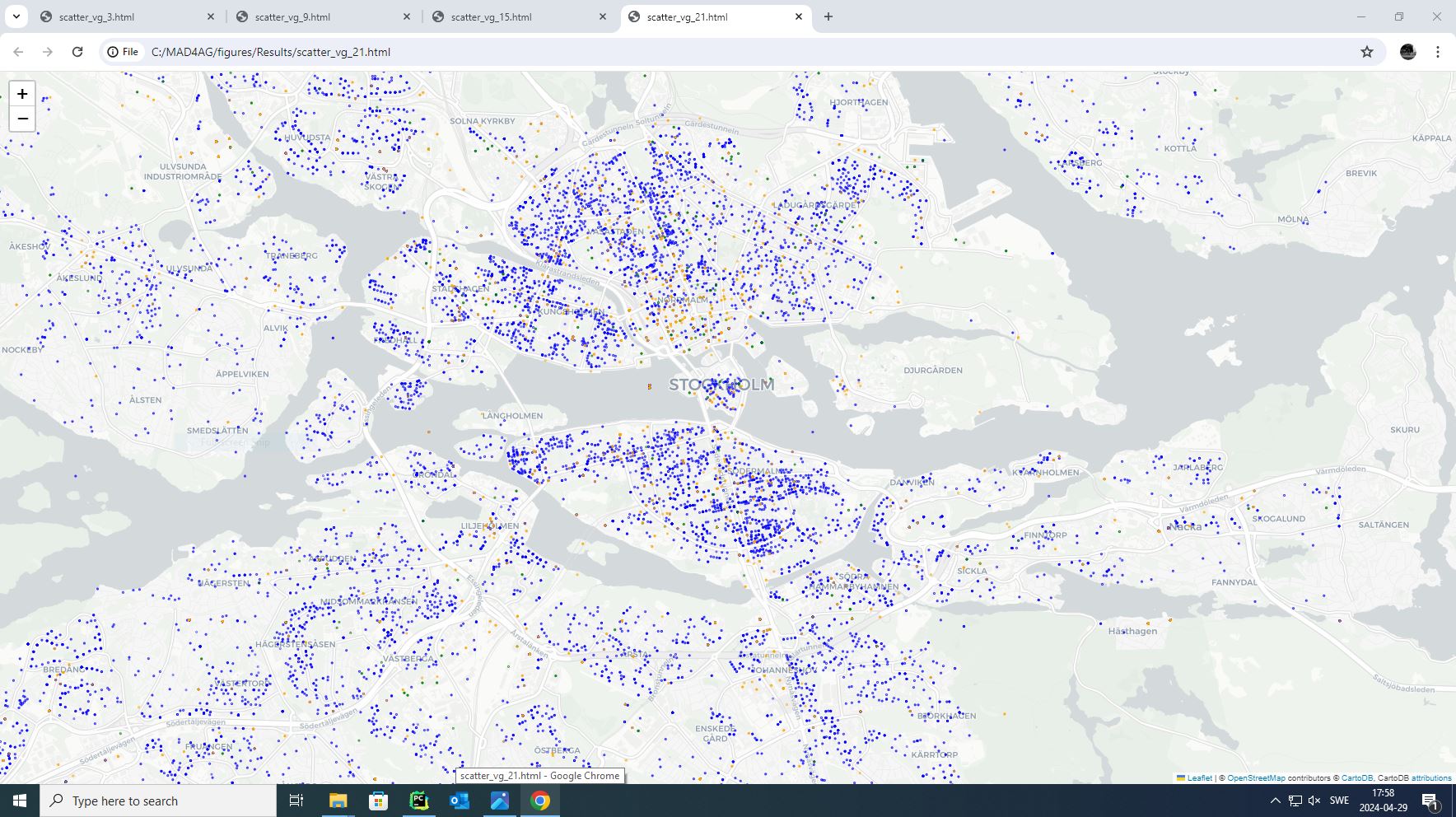} \\
            09:00 pm \\[6pt]
        
        \end{minipage}
    \end{tabular}
    \caption{Distribution of activities by time of day in central Stockholm.}
    \label{fig:geolocationsbyactivities}
\end{figure}

Figure \ref{fig:geolocationsbyactivities} illustrates the distribution of activity participation by activity type and time of day in central Stockholm. The home activity locations are distributed across Stockholm, and the activity participation occurs throughout the day. The work activities predominantly occur during daylight hours. The other activities are widespread on the map and show increased intensity during the daytime, with some extending into the night hours.

\end{document}